\documentclass[ aps, twocolumn, showkeys, notitlepage, superscriptaddress]{revtex4-2}
 
\usepackage{natbib}
\usepackage{lipsum}
\usepackage{graphicx}
\usepackage{dcolumn}
\usepackage{bm}
\usepackage{amsmath}
\usepackage{float}
\usepackage{multirow}
\usepackage{slashed}
\usepackage{xcolor}
\usepackage{physics}
\usepackage{multirow}
\usepackage{gensymb}
\usepackage[colorlinks=true, pdfstartview=FitV, bookmarks=true, bookmarksnumbered=true, breaklinks]{hyperref}
\usepackage{mathtools,braket}
\usepackage{lipsum}  
\usepackage{color}
\definecolor{blue}{rgb}{0.0, 0.0, 1.0}
\definecolor{red}{rgb}{1.0, 0.0, 0.0}
\definecolor{royalblue}{rgb}{0.0, 0.14, 0.4}
\hypersetup{linkcolor=blue, citecolor=blue, urlcolor=blue}

\def\orcid#1{\kern .08em\href{https://orcid.org/#1}{\includegraphics[keepaspectratio,width=0.7em]{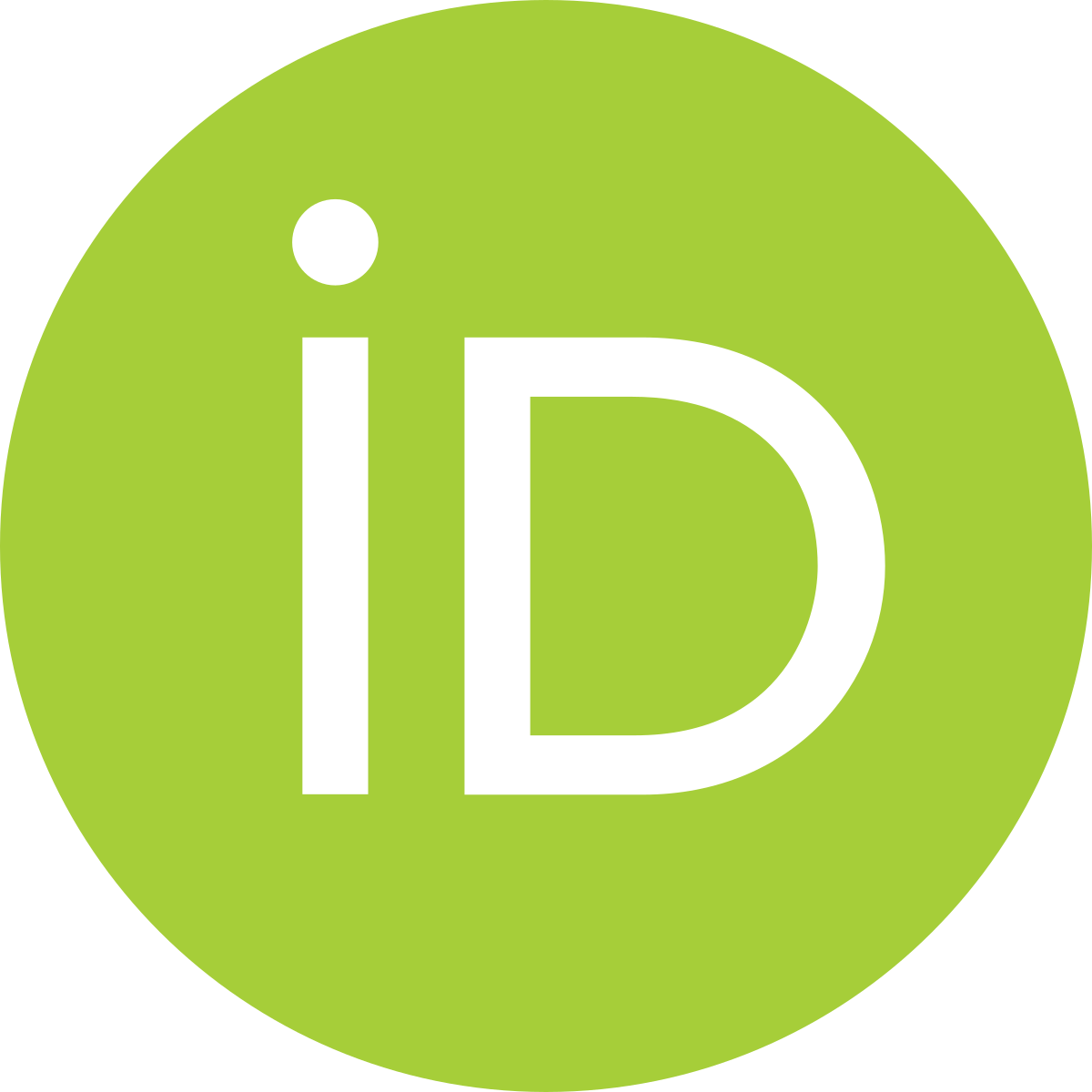}}}

\begin{document}

\title{Quarkonium spectra with magnetically induced anisotropic confinement}

\author{Ahmad Jafar Arifi\orcid{0000-0002-9530-8993}} 
\email{aj.arifi01@gmail.com}
\affiliation{Advanced Science Research Center, Japan Atomic Energy Agency, Tokai, Ibaraki 319-1195, Japan}
\affiliation{Research Center for Nuclear Physics, The University of Osaka, Ibaraki, Osaka 567-0047, Japan}

\author{Kei Suzuki\orcid{0000-0002-8746-4064}}
\email{k.suzuki.2010@th.phys.titech.ac.jp}
\affiliation{Advanced Science Research Center, Japan Atomic Energy Agency, Tokai, Ibaraki 319-1195, Japan}

\date{\today}

\begin{abstract} 
Strong magnetic fields modify the force that confines quarks inside hadrons and make it direction-dependent.
Using quark-antiquark potentials obtained from lattice simulations as inputs to a quark potential model, we investigate how the anisotropic confinement affects the mass spectrum of quarkonium.
In the strong-field regime, we find downward mass shifts induced by a softening of the confining potential along the field direction.
In particular, the mass shifts of radially excited states are more significant than that of the ground state. For the longitudinal spin eigenstates, the excited-state spectrum strongly depends on the magnetic-field strength, in contrast to the spectrum with conventional isotropic confinement, which is insensitive to the field strength. 
This provides a clean probe of magnetically induced confinement anisotropy that can be confirmed in future lattice simulations.
\end{abstract}

\maketitle

\section{Introduction}

Understanding how fundamental interactions respond to external influences is a recurring theme across many areas of physics.
In strongly interacting systems, such external influences are particularly informative, as they offer new opportunities to explore the formation of bound states and the response of their internal dynamics to changes in the surrounding environment~\cite{Hattori:2016emy,Zhao:2020jqu,Iwasaki:2021nrz}.
Quantum chromodynamics (QCD), the theory of the strong interaction, remains not fully understood under external conditions, making it a compelling subject for investigation.
Because QCD is inherently nonperturbative at hadronic scales, first-principles numerical approaches, i.e., lattice QCD simulations, play a central role in revealing the mechanisms of confinement~\cite{Bonati:2014ksa,Bonati:2016kxj,DElia:2021tfb} and hadron properties~\cite{Bali:2011qj,Luschevskaya:2012xd,Hidaka:2012mz,Luschevskaya:2014lga,Luschevskaya:2015cko,Bali:2017ian,Bali:2018sey,Luschevskaya:2018chr,Hattori:2019ijy,Bignell:2019vpy,Bignell:2020dze,Ding:2020hxw,Ding:2022tqn,Endrodi:2024cqn,Ding:2025pbu,Ding:2026qzu} in magnetic fields.
Moreover, strong magnetic fields are generally relevant to heavy-ion collisions, where transient fields up to $eB \sim 1~\mathrm{GeV}^2$ may be generated in the early stages of collisions~\cite{Kharzeev:2007jp,Skokov:2009qp,Voronyuk:2011jd,Deng:2012pc}.

\begin{figure*}[t]
    \centering
    \includegraphics[width=1.4\columnwidth]{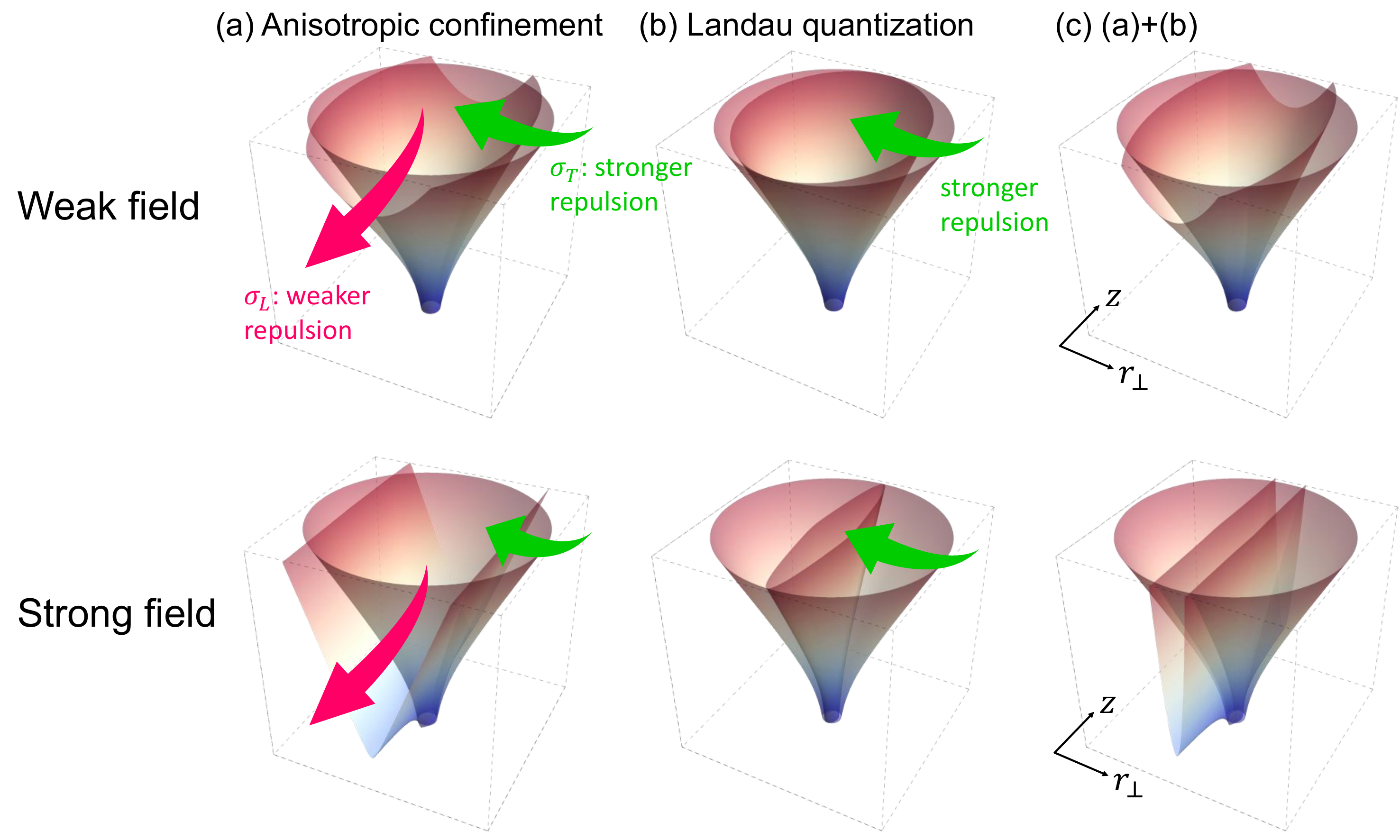} 
\caption{Schematics of magnetic-field-induced anisotropy effects in quarkonia in the weak- and strong-field regimes. (a) Anisotropic confinement potential, (b) harmonic-oscillator potential arising from Landau quantization of quarks, and (c) the sum of both contributions. The conventional isotropic (i.e., linear and Coulomb) potential is also plotted for comparison.}
    \label{fig:schematic}
\end{figure*}

Lattice QCD simulations have shown that a strong magnetic field induces a pronounced anisotropy in the quark-antiquark potential~\cite{Bonati:2014ksa,Bonati:2016kxj}, which has recently been explored up to \(eB = 9~\mathrm{GeV}^2\)~\cite{DElia:2021tfb}.
In particular, the confining force becomes stronger transverse to the magnetic field and weaker along it, as illustrated schematically in Fig.~\hyperref[fig:schematic]{\ref*{fig:schematic}(a)}, while the accompanying anisotropy of the short-range Coulomb interaction was found to be small~\cite{Bonati:2016kxj}.
This naturally raises the question of how hadronic bound states respond when the underlying interaction becomes anisotropic. 

Heavy quarkonia provide a particularly clean system for addressing this problem.
As nonrelativistic bound states characterized by well-separated energy scales, their spectra are tightly governed by the shape of the interquark potential~\cite{Eichten:1979ms,Buchmuller:1980su,Godfrey:1985xj}.
Radially excited states, which probe larger spatial regions, are therefore especially sensitive to anisotropic deformations of the confining interaction, making them powerful probes of directional modifications induced by strong magnetic fields.

The properties of quarkonia in external magnetic fields, both static~\cite{Marasinghe:2011bt,Tuchin:2011cg,Yang:2011cz,Tuchin:2013ie,Machado:2013rta,Alford:2013jva,Cho:2014exa,Dudal:2014jfa,Cho:2014loa,Bonati:2015dka,Sadofyev:2015hxa,Suzuki:2016kcs,Yoshida:2016xgm,Hasan:2017fmf,Singh:2017nfa,Braga:2018zlu,Iwasaki:2018pby,Amal:2018qln,Iwasaki:2018czv,Hasan:2018kvx,Braga:2019yeh,Hasan:2020iwa,Zhou:2020ssi,Braga:2020hhs,Braga:2021fey,Jena:2022nzw,Hu:2022ofv,Ghosh:2022sxi,Parui:2022msu,Sebastian:2023tlw,Nilima:2024nvd,Jena:2024cqs,Shukla:2024qlf,Wen:2025dwy,Jena:2025xcf,Arifi:2025ivt,Dominguez:2025nar}
and dynamical~\cite{Guo:2015nsa,Suzuki:2016fof,Dutta:2017pya,Hoelck:2017dby,Bagchi:2018olp,Iwasaki:2021kms,Arifi:2025atv}, have been extensively investigated. However, the specific impact of magnetically induced anisotropy of the confining potential itself has received comparatively less attention~\cite{Miransky:2002rp,Chernodub:2010xce,Andreichikov:2012xe,Bonati:2015dka,Simonov:2015yka,Bohra:2019ebj,Gursoy:2020kjd,Arefeva:2020bjk,Arefeva:2026yms}.

Motivated by recent lattice QCD results on anisotropic confinement~\cite{Bonati:2014ksa,Bonati:2016kxj,DElia:2021tfb}, in this work we explore how such anisotropy manifests itself in the quarkonium spectrum.
By employing anisotropic string tensions as inputs to a quark potential model~\cite{Suzuki:2016kcs,Yoshida:2016xgm}, we find substantial and robust modifications of the level spacing, with particularly large downward shifts of radially excited states driven primarily by the weakened longitudinal confinement.
While earlier quark-model study~\cite{Bonati:2015dka} focused mainly on the weak-field regime and radial ground states, where anisotropic Coulomb effects play a more prominent role, more recent lattice QCD results~\cite{DElia:2021tfb} enable a systematic investigation of spectral modifications in the strong-field regime.
In this region, excited quarkonium states provide especially sensitive probes of anisotropic confinement and offer clear insight into the directional restructuring of the strong interaction under extreme conditions.

\section{Model}
We first present the calculation of quarkonium spectra within a simple quark model under a static and uniform magnetic field $\bm{B} = B \hat{z}$, taking charmonium as a representative example.
The Hamiltonian for the two-body system consisting of charm and anticharm quarks is~\cite{Alford:2013jva}
\begin{equation}
    H_{c\bar{c}} = \sum_{i=c,\bar{c}} \left[
        m_i + \frac{(\bm{p}_i - q_i \bm{A})^2}{2m_i}
        - \bm{\mu}_i \cdot \bm{B}
    \right]
    + V_{c\bar{c}}(r),
\end{equation}
where $m_i$, $\bm{p}_i$, and $q_i$ are the quark mass, momentum, and electric charge, respectively.  
We adopt the symmetric gauge $\bm{A} = \frac12 \bm{B}\times \bm{r}$ and use the quark magnetic moment
$\bm{\mu}_i = g q_i \bm{S}_i/(2m_i)$, where the Land{\'e g-}factor $g=2$ and $q_c= -q_{\bar{c}}=(2/3)e$ with the elementary charge $e$.

The quark-antiquark interaction is modeled as
\begin{equation}
V_{c\bar{c}}(r)
= C + \sigma r - \frac{\alpha}{r}
+ \beta(\bm{S}_c\!\cdot\!\bm{S}_{\bar c}) e^{-\Lambda r^2},
\end{equation}
where $r=\sqrt{z^2+r_\perp^2}$ is the relative distance.  
The parameters $\sigma$, $\alpha$, and $\beta$ represent the confining, Coulomb, and spin-spin interactions, respectively, and $\Lambda$ controls Gaussian smearing of the spin-spin potential~\cite{Barnes:2005pb}.  
The constant $C$ fixes the overall mass shift. 
For the model parameters, we use the same set as Refs.~\cite{Suzuki:2016kcs,Yoshida:2016xgm}, where $m_c=1.784\,{\rm GeV}$, $\alpha=0.713$, $\sqrt{\sigma}=0.402\,{\rm GeV}$, and
 $\Lambda=1.020\,{\rm GeV}^2$ were determined from lattice QCD simulations~\cite{Kawanai:2015tga,Kawanai:2011jt}, while the remaining parameters, $\beta=0.4778\,{\rm GeV}$ and $C=-0.5693\,{\rm GeV}$ were fixed to reproduce the experimental masses of the ground-state $\eta_c$ and $J/\psi$.

In a uniform magnetic field, the ordinary translational symmetry is broken, but a generalized translational symmetry is preserved, leading to the conservation of pseudomomentum~\cite{Johnson:1949} $\bm{K}=\sum_{i=1}^2\left[\bm{p}_i + \frac{1}{2}q_i\,\bm{B}\times\bm{r}_i\right]$, instead of the canonical center-of-mass momentum. In the present work, we focus on the case of $\bm{K}=0$, where the center-of-mass and relative motions decouple. Consequently, for charmonia, i.e., equal-mass quark-antiquark systems with opposite electric charges, the total wave function can be factorized as $\Phi(\bm{R},\bm{r})=\exp\left[i\left(\bm{K}-\frac{1}{2}q\,\bm{B}\times \bm{r}\right)\cdot \bm{R}\right]\psi(\bm{r})$, and the Hamiltonian reduces to the relative form~\cite{Machado:2013rta,Alford:2013jva,Andreichikov:2013zba}
\begin{equation}
\label{eq:H_rel}
    H_{c\bar{c}} = -\frac{\bm{\nabla}^2}{2\mu}
    + \frac{q_c^2B^2}{8\mu} r_\perp^2
    + V_{c\bar{c}}(r)
    - \sum_{i=c,\bar c} \bm{\mu}_i\cdot\bm{B},
\end{equation}
where $\mu=m_c/2$.  
The $B^2 r_\perp^2$ term represents the quark-Landau-level contribution and preserves cylindrical symmetry, as illustrated in Fig.~\hyperref[fig:schematic]{\ref*{fig:schematic}(b)}.

The last term in Eq.~(\ref{eq:H_rel}), the quark Zeeman interaction, leads to mixing between the pseudoscalar quarkonia $\ket{00}$ and the longitudinal spin component $\ket{10}$ of vector quarkonia, where the state is labeled by $\ket{SS_z}$ with total spin $S$ and its third component $S_z$, while the transverse components $\ket{S_z=\pm 1}$ remain unaffected by this mixing. This is because, for equal-mass quarkonia such as charmonium, the magnetic-moment contributions from the quark and antiquark cancel in the transverse channel, resulting in a vanishing net Zeeman shift~\cite{Machado:2013rta, Alford:2013jva,Yoshida:2016xgm}. The surviving off-diagonal matrix element is
\begin{equation}\label{eq:zeeman}
\bra{10}\! -(\bm{\mu}_c+\bm{\mu}_{\bar{c}})\cdot\bm{B} \!\ket{00}
= -\,\frac{g q B}{4\mu},
\end{equation}
which generates a coupled-channel Schrödinger equation for the $S_z=0$ sector. Due to this mixing, the longitudinal states are labeled as the 1st, 2nd, 3rd, 4th, etc., while the transverse states are denoted as $J/\psi_T, \psi_T^\prime$, and so on. We then solve the Schrödinger equation using the cylindrical Gaussian expansion method (CGEM)~\cite{Suzuki:2016kcs,Yoshida:2016xgm}, an extension of the Gaussian expansion method~\cite{Kamimura:1988zz,Hiyama:2003cu} optimized for systems with cylindrical symmetry, allowing an accurate determination of spectra and wave functions in external magnetic fields.

\begin{figure}[b]
    \centering
    \includegraphics[width=0.95\columnwidth]{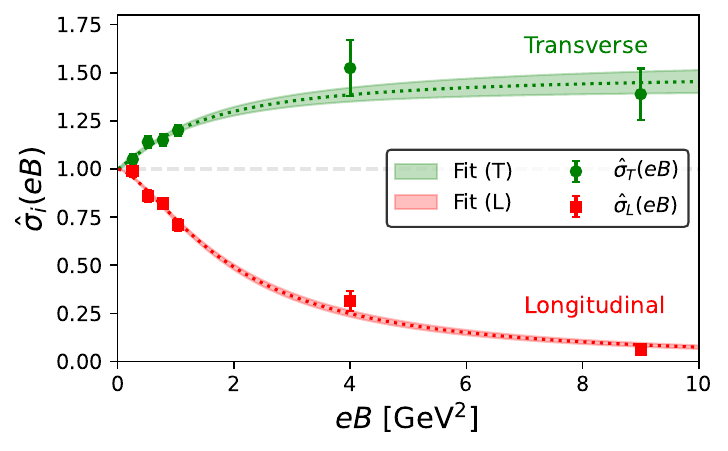}
\caption{Magnetic-field-induced anisotropy of the string tension, extracted from our fits and compared with available lattice-QCD results~\cite{Bonati:2016kxj,DElia:2021tfb}, where we put the four data points at the finest lattice spacing $a = 0.0989$ fm in Ref.~\cite{Bonati:2016kxj} and the two data at $eB=4,9$ GeV$^2$ in Ref.~\cite{DElia:2021tfb}. 
}
    \label{fig:anisotropic}
\end{figure}

\begin{figure*}[t]
    \centering 
    \includegraphics[width=2.0\columnwidth]{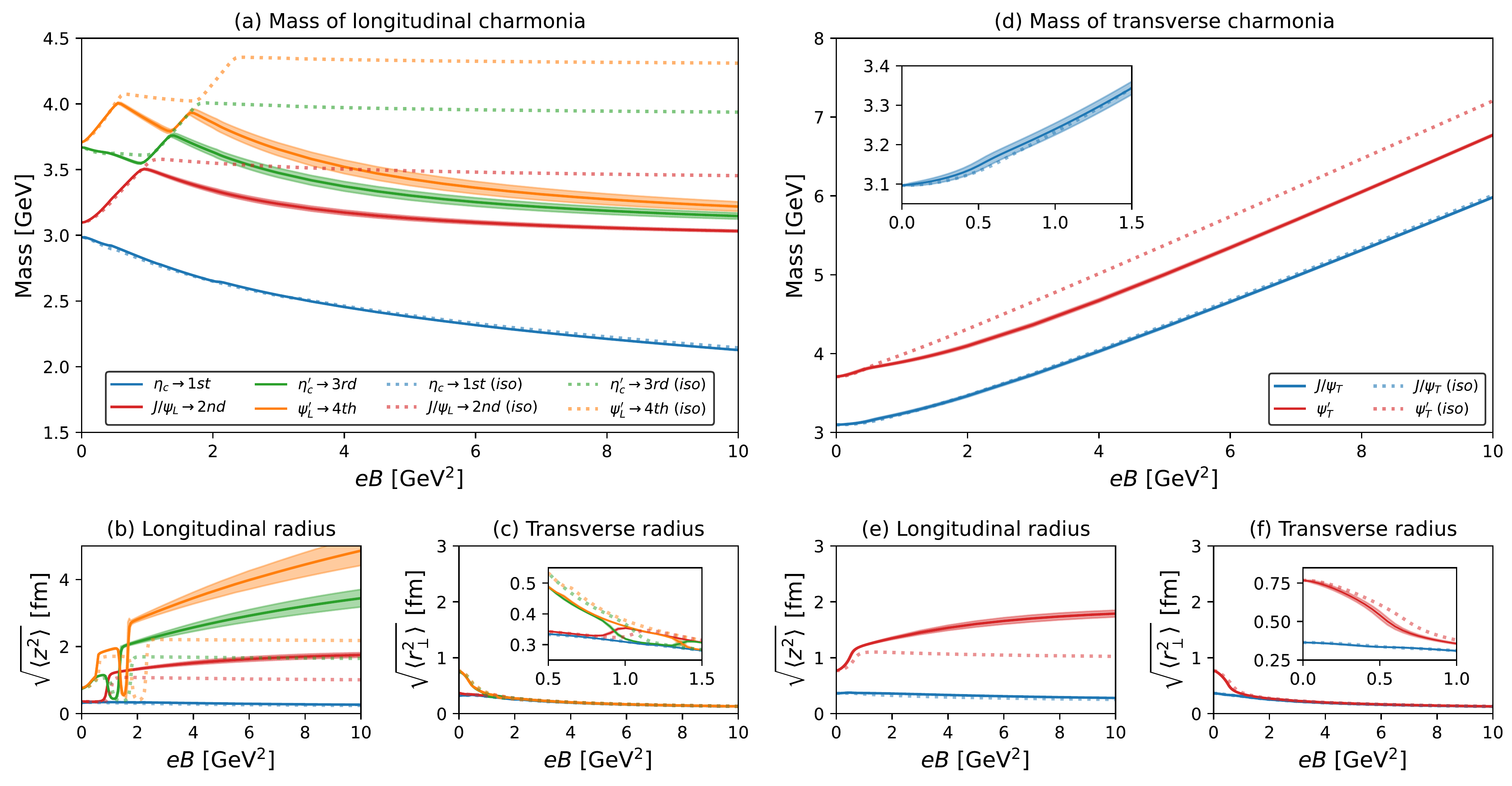}
    \caption{ Charmonium masses and root-mean-square radii as functions of the magnetic-field strength $eB$ under anisotropic confinement. The left and right panels correspond to longitudinal ($S_z = 0$) and transverse ($S_z = \pm 1$) charmonia, respectively. Panels (a) and (d) show the mass spectra, while panels (b), (c), (e), and (f) display the longitudinal and transverse root-mean-square radii, $\sqrt{\langle z^2 \rangle}$ and $\sqrt{\langle r_\perp^2 \rangle}$, respectively. Solid lines show the results obtained with anisotropic confinement. Dotted lines show the corresponding results with isotropic confinement. Shaded bands show the uncertainties propagated from the fitting uncertainties of the anisotropic string tensions.
    } 
    \label{fig:result}
\end{figure*}

\section{Anisotropy}

Lattice QCD simulations show that strong magnetic fields deform the string tension, where the transverse confinement is enhanced while the longitudinal one is suppressed~\cite{Bonati:2014ksa,Bonati:2016kxj,DElia:2021tfb}.
Since lattice-QCD simulations provide the string tensions along the longitudinal and transverse directions, we adopt the simplest smooth three-dimensional interpolation consistent with cylindrical symmetry around the magnetic-field axis~\cite{Bonati:2015dka},
\begin{equation}
\label{eq:V_aniso}
V_{\rm conf}^{\rm aniso}(r_\perp,z)
= \sigma(0)\sqrt{\epsilon_L z^2 + \epsilon_T r_\perp^2 },
\end{equation}
which correctly reproduces the lattice-QCD string tensions in the longitudinal and transverse limits. 
Here, $\epsilon_{T,L}$ are related to the transverse and longitudinal string tensions via $\sigma_{T,L}/\sigma(0)=\sqrt{\epsilon_{T,L}}$. 
The magnetic-field dependence of the normalized string tensions,
$\hat{\sigma}_i(eB)={\sigma_i(eB)}/{\sigma(0)},$
is parametrized as
\begin{equation}
\hat{\sigma}_{i}(eB)
= a_i + \frac{1 - a_i}{1 + \kappa_{i}(eB)^{\gamma_{i}}},
\qquad i=T,L,
\end{equation}
which exhibits saturation behavior $\hat{\sigma}_{i}(eB\to \infty)\to a_i.$
In particular, we impose that the longitudinal string tension vanishes in this limit, corresponding to $a_L=0$. Moreover, although uncertainties for the transverse component are much larger, we find that its effect is insignificant and is effectively screened by Landau quantization.
Here, $\kappa_i$ determines the scale of the magnetic-field dependence, while $\gamma_i$ controls the steepness of the transition. 
By fitting to the lattice-QCD data, we obtain
\begin{align}
\{a_T,\kappa_T,\gamma_T\} &= \{\;1.50(9),\; 0.66(22),\; 1.19(22)\;\}, \\
\{a_L,\kappa_L,\gamma_L\} &= \{\;0.00,\; 0.356(18),\; 1.543(58)\;\},
\end{align}
with $\chi_T^2/\mathrm{d.o.f.}=0.68,$ and $\chi_L^2/\mathrm{d.o.f.}=1.17,$
for the transverse and longitudinal cases, respectively
\footnote{We note that the transverse-sector parameters exhibit strong correlations, particularly between $a_T$ and $\kappa_T$, since the asymptotic region is less constrained, whereas the longitudinal-sector parameters are only weakly correlated because we impose $a_L=0$.}. 
Figure~\ref{fig:anisotropic} shows that this parametrization well reproduces the lattice-QCD data.

\section{Spectroscopy}

Upper panels of Fig.~\ref{fig:result} demonstrate that the longitudinal ($S_z=0$) and transverse ($S_z=\pm1$) eigenstates of charmonia respond in qualitatively different ways to anisotropic confinement in strong magnetic fields, revealing a striking pattern compared with the isotropic case.
We find that pronounced downward mass shifts develop for excited charmonia, while the ground state is relatively insensitive.
As a result, the avoided crossings between excited states are shifted toward smaller $eB$ values. In particular, the avoided-crossing positions move from $eB \approx 1.79$ to $1.38~\mathrm{GeV}^2$ for the 4th--3rd states and from $1.08$ to $0.93~\mathrm{GeV}^2$ for the 3rd--2nd states.
This shift of the level repulsion structure toward lower magnetic fields has important implications for determining the nonadiabatic transition rates in transient magnetic fields~\cite{Arifi:2025atv}.

The downward mass shifts for the ground and excited states originate from the reduction of the longitudinal string tension~\cite{DElia:2021tfb}, which weakens confinement along the magnetic-field direction and extends the wave function longitudinally, thereby lowering the excitation energies.
At the same time, transverse squeezing plays a minor role because its effect is almost screened by the freezing of transverse motion into the lowest Landau level (LLL), as illustrated in Fig.~\hyperref[fig:schematic]{\ref*{fig:schematic}(c)}. However, in the weak-field regime, where the Landau quantization is less significant, the mass increases due to the enhancement of transverse confinement, although this effect is rather small, as shown in the inset of Fig.~\hyperref[fig:result]{\ref*{fig:result}(d)}.
 
This behavior contrasts with the isotropic potential case~\cite{Suzuki:2016kcs,Yoshida:2016xgm} shown by the dotted lines in Fig.~\hyperref[fig:result]{\ref*{fig:result}(a)}.
In that scenario, the masses of excited longitudinal charmonia saturate at large $eB$ once the transverse motion is frozen into the LLL and spin mixing [Eq.~\eqref{eq:zeeman}] reaches its asymptotic limit. 
Then, due to the cancellation between the positive $eB$ dependence of the LLL energy and the negative Zeeman shift, the spectrum in the strong-$eB$ regime is flattened into a constant plateau, whereas the masses of transverse charmonia continue to increase due to the LLL energy in the absence of Zeeman shifts. 
Note that such a flattening mechanism is also expected to hold for the ground state; however, it has not yet saturated for charmonium in the current plot range, while this saturation can be seen in the light-meson sector~\cite{Orlovsky:2013gha,Taya:2014nha,Andreichikov:2016ayj,Kojo:2021gvm}.

Although our calculation is based on a nonrelativistic framework, the same saturation mechanism holds even in the relativistic picture.
The relativistic Landau levels for Dirac fermions are represented as~\cite{Rabi:1928,Johnson:1949}
\begin{equation}
E_{n,s} = \sqrt{p_z^2 + m^2 + |qB|(2n+1-s)}.
\end{equation}
This form contains the Zeeman shift through the spin index $s$ and implies that the relativistic LLL ($n=0$, $s=+1$) is independent of the magnetic field.
Since quarkonia in the $B\to\infty$ limit are dominated by the LLLs of the quark and antiquark, the saturated energy can be written as
\begin{equation}
E_{\parallel}(B\to\infty)
\approx
\sqrt{p_{z,c}^2+m^2_c}
+
\sqrt{p_{z,\bar{c}}^2+m^2_{\bar{c}}}.
\end{equation}
For discussion on neutral light mesons, also see Refs.~\cite{Orlovsky:2013gha,Taya:2014nha,Andreichikov:2016ayj,Kojo:2021gvm}.  

The anisotropic confinement case is therefore qualitatively distinct: because the longitudinal potential itself is modified, the spectrum continues to evolve even in regimes where isotropic systems have already saturated. This behavior is further illustrated in Table~\ref{tab:splitting}. In particular, in the anisotropic case, higher-state splittings decrease significantly with increasing magnetic field: the $3\mathrm{rd}-2\mathrm{nd}$ and $4\mathrm{th}-3\mathrm{rd}$ splitting is reduced by about $122$ and $106~\mathrm{MeV}$, respectively, from $eB=3$ to $10~\mathrm{GeV}^2$. In contrast, the splittings in the isotropic case remain nearly constant, showing only a slight increase, which indicates a much more stable excited-state spectrum.

It is worth noting that other anisotropies are in principle also possible, such as an anisotropic Coulomb potential~\cite{Bonati:2014ksa,Bonati:2015dka}.  
However, more recent lattice QCD results indicate that the anisotropy of the Coulomb term is relatively small~\cite{Bonati:2016kxj}. Since the Coulomb term mainly governs the short-range part of the potential, its effects are expected to be more pronounced for the ground states, while the impact on excited states should remain tiny compared with that from the anisotropic confinement.

\begin{table}[b]
\caption{Mass splittings between adjacent states for the anisotropic (Aniso.) and isotropic (Iso.) cases at different magnetic fields $eB$. The numbers in parentheses denote the uncertainties in the anisotropic results.}
\label{tab:splitting}
\begin{ruledtabular}
\begin{tabular}{c cc cc cc}
& \multicolumn{2}{c}{$2\mathrm{nd}-1\mathrm{st}$}
& \multicolumn{2}{c}{$3\mathrm{rd}-2\mathrm{nd}$}
& \multicolumn{2}{c}{$4\mathrm{th}-3\mathrm{rd}$} \\
$eB$ & Aniso. & Iso. & Aniso. & Iso. & Aniso. & Iso. \\
$[\mathrm{GeV}^2]$ & [GeV] & [GeV]& [GeV]& [GeV]& [GeV]& [GeV]\\
\hline
0  & 0.111 & 0.111 & 0.572 & 0.572 & 0.038 & 0.038 \\
3  & 0.694(10) & 0.982 & 0.236(20) & 0.464 & 0.179(31) & 0.363 \\
5  & 0.749(8) & 1.101 & 0.174(19) & 0.471 & 0.126(31) & 0.368 \\
7  & 0.814(6) & 1.197 & 0.141(16) & 0.477 & 0.097(27) & 0.370 \\
9  & 0.876(5) & 1.275 & 0.121(14) & 0.482 & 0.079(23) & 0.372 \\
10 & 0.905(4) & 1.310 & 0.114(12) & 0.483 & 0.073(22) & 0.373 \\
\end{tabular}
\end{ruledtabular}
\end{table}

\section{Deformation}

Lower panels of Fig.~\ref{fig:result} illustrate the magnetic-field dependence of charmonium radii, defined as $\sqrt{\langle z^{2}\rangle}\equiv\sqrt{3\langle \psi|z^{2}|\psi\rangle}$ and $\sqrt{\langle r_\perp^{2}\rangle}\equiv\sqrt{(3/2)\langle \psi|r_\perp^{2}|\psi\rangle}$~\cite{Suzuki:2016kcs,Yoshida:2016xgm}.
In the isotropic case (dotted lines), the transverse radius is strongly suppressed by Landau-level squeezing, while the longitudinal radius remains nearly unchanged at large $eB$. This behavior reflects that the transverse motions of quarks are frozen into the LLL, while the longitudinal motion is not affected.

When anisotropic confinement [Eq.~\eqref{eq:V_aniso}] is introduced, the radii of the excited states undergo substantial modifications, as shown in Fig.~\hyperref[fig:result]{\ref*{fig:result}(b)}. The magnetic field significantly weakens the longitudinal confinement, producing a much shallower potential along the field direction. As a result, the quarkonium wave function becomes strongly elongated longitudinally, leading to a large increase in $\sqrt{\langle z^{2}\rangle}$. This elongation is more pronounced for higher radial excitations, since their larger size is more sensitive to the weakened confinement.  

In strong magnetic fields, although the transverse confinement is enhanced, its effect remains subdominant compared with the Landau-level squeezing, and the transverse radius $\sqrt{\langle r_\perp^{2}\rangle}$ is therefore nearly unaffected by the anisotropic confinement.
In smaller magnetic fields, the strengthened transverse confinement becomes more visible, leading to a slight reduction of the transverse radius, as shown in Fig.~\hyperref[fig:result]{\ref*{fig:result}(c)}.

\section{Conclusion}

In summary, we have explored how the lattice-QCD-motivated anisotropic confinement potential induced by strong magnetic fields~\cite{Bonati:2014ksa,Bonati:2016kxj,DElia:2021tfb} modifies the mass spectra and structure of charmonia by precisely solving a quark potential model.
We find that the dominant effect originates from the substantial weakening of the longitudinal string tension, which leads to pronounced downward mass shifts of radially excited longitudinal charmonia, in sharp contrast to the characteristic mass plateau observed in the isotropic case in the strong magnetic-field regime.
By comparison, the ground state is only mildly affected, which emphasizes the sensitivity of excited states to the anisotropic confinement.
This reduced longitudinal confinement also induces a strong elongation of the wave function along the magnetic-field direction, as characterized by a significant increase in the longitudinal radius $\sqrt{\langle z^{2}\rangle}$, whereas in the isotropic case the radius saturates.
Thus, our results establish a transparent and quantitative connection between lattice-QCD results and quarkonium structure in extreme magnetic environments.

\section*{Acknowledgments}

A. J. A. was supported by the JAEA Postdoctoral Fellowship Program, and partially by the RCNP Collaboration Research Network Program under Project No. COREnet 057, as well as by the PUTI Q1 Grant from the University of Indonesia under Contract No. PKS-206/UN2.RST/HKP.05.00/2025.
This work was supported by the Japan Society for the Promotion of Science (JSPS) KAKENHI (Grant No. JP24K07034 for K. S.) and (Grant No. JP26K17163 for A.J.A.).

\bibliography{references}

\begin{thebibliography}{94}%
\makeatletter
\providecommand \@ifxundefined [1]{%
 \@ifx{#1\undefined}
}%
\providecommand \@ifnum [1]{%
 \ifnum #1\expandafter \@firstoftwo
 \else \expandafter \@secondoftwo
 \fi
}%
\providecommand \@ifx [1]{%
 \ifx #1\expandafter \@firstoftwo
 \else \expandafter \@secondoftwo
 \fi
}%
\providecommand \natexlab [1]{#1}%
\providecommand \enquote  [1]{``#1''}%
\providecommand \bibnamefont  [1]{#1}%
\providecommand \bibfnamefont [1]{#1}%
\providecommand \citenamefont [1]{#1}%
\providecommand \href@noop [0]{\@secondoftwo}%
\providecommand \href [0]{\begingroup \@sanitize@url \@href}%
\providecommand \@href[1]{\@@startlink{#1}\@@href}%
\providecommand \@@href[1]{\endgroup#1\@@endlink}%
\providecommand \@sanitize@url [0]{\catcode `\\12\catcode `\$12\catcode `\&12\catcode `\#12\catcode `\^12\catcode `\_12\catcode `\%12\relax}%
\providecommand \@@startlink[1]{}%
\providecommand \@@endlink[0]{}%
\providecommand \url  [0]{\begingroup\@sanitize@url \@url }%
\providecommand \@url [1]{\endgroup\@href {#1}{\urlprefix }}%
\providecommand \urlprefix  [0]{URL }%
\providecommand \Eprint [0]{\href }%
\providecommand \doibase [0]{https://doi.org/}%
\providecommand \selectlanguage [0]{\@gobble}%
\providecommand \bibinfo  [0]{\@secondoftwo}%
\providecommand \bibfield  [0]{\@secondoftwo}%
\providecommand \translation [1]{[#1]}%
\providecommand \BibitemOpen [0]{}%
\providecommand \bibitemStop [0]{}%
\providecommand \bibitemNoStop [0]{.\EOS\space}%
\providecommand \EOS [0]{\spacefactor3000\relax}%
\providecommand \BibitemShut  [1]{\csname bibitem#1\endcsname}%
\let\auto@bib@innerbib\@empty
\bibitem [{\citenamefont {Hattori}\ and\ \citenamefont {Huang}(2017)}]{Hattori:2016emy}%
  \BibitemOpen
  \bibfield  {author} {\bibinfo {author} {\bibfnamefont {K.}~\bibnamefont {Hattori}}\ and\ \bibinfo {author} {\bibfnamefont {X.-G.}\ \bibnamefont {Huang}},\ }\bibfield  {title} {\bibinfo {title} {{Novel quantum phenomena induced by strong magnetic fields in heavy-ion collisions}},\ }\href {https://doi.org/10.1007/s41365-016-0178-3} {\bibfield  {journal} {\bibinfo  {journal} {Nucl. Sci. Tech.}\ }\textbf {\bibinfo {volume} {28}},\ \bibinfo {pages} {26} (\bibinfo {year} {2017})},\ \Eprint {https://arxiv.org/abs/1609.00747} {arXiv:1609.00747 [nucl-th]} \BibitemShut {NoStop}%
\bibitem [{\citenamefont {Zhao}\ \emph {et~al.}(2020)\citenamefont {Zhao}, \citenamefont {Zhou}, \citenamefont {Chen},\ and\ \citenamefont {Zhuang}}]{Zhao:2020jqu}%
  \BibitemOpen
  \bibfield  {author} {\bibinfo {author} {\bibfnamefont {J.}~\bibnamefont {Zhao}}, \bibinfo {author} {\bibfnamefont {K.}~\bibnamefont {Zhou}}, \bibinfo {author} {\bibfnamefont {S.}~\bibnamefont {Chen}},\ and\ \bibinfo {author} {\bibfnamefont {P.}~\bibnamefont {Zhuang}},\ }\bibfield  {title} {\bibinfo {title} {{Heavy flavors under extreme conditions in high energy nuclear collisions}},\ }\href {https://doi.org/10.1016/j.ppnp.2020.103801} {\bibfield  {journal} {\bibinfo  {journal} {Prog. Part. Nucl. Phys.}\ }\textbf {\bibinfo {volume} {114}},\ \bibinfo {pages} {103801} (\bibinfo {year} {2020})},\ \Eprint {https://arxiv.org/abs/2005.08277} {arXiv:2005.08277 [nucl-th]} \BibitemShut {NoStop}%
\bibitem [{\citenamefont {Iwasaki}\ \emph {et~al.}(2021{\natexlab{a}})\citenamefont {Iwasaki}, \citenamefont {Oka},\ and\ \citenamefont {Suzuki}}]{Iwasaki:2021nrz}%
  \BibitemOpen
  \bibfield  {author} {\bibinfo {author} {\bibfnamefont {S.}~\bibnamefont {Iwasaki}}, \bibinfo {author} {\bibfnamefont {M.}~\bibnamefont {Oka}},\ and\ \bibinfo {author} {\bibfnamefont {K.}~\bibnamefont {Suzuki}},\ }\bibfield  {title} {\bibinfo {title} {{A review of quarkonia under strong magnetic fields}},\ }\href {https://doi.org/10.1140/epja/s10050-021-00533-5} {\bibfield  {journal} {\bibinfo  {journal} {Eur. Phys. J. A}\ }\textbf {\bibinfo {volume} {57}},\ \bibinfo {pages} {222} (\bibinfo {year} {2021}{\natexlab{a}})},\ \Eprint {https://arxiv.org/abs/2104.13990} {arXiv:2104.13990 [hep-ph]} \BibitemShut {NoStop}%
\bibitem [{\citenamefont {Bonati}\ \emph {et~al.}(2014)\citenamefont {Bonati}, \citenamefont {D'Elia}, \citenamefont {Mariti}, \citenamefont {Mesiti}, \citenamefont {Negro},\ and\ \citenamefont {Sanfilippo}}]{Bonati:2014ksa}%
  \BibitemOpen
  \bibfield  {author} {\bibinfo {author} {\bibfnamefont {C.}~\bibnamefont {Bonati}}, \bibinfo {author} {\bibfnamefont {M.}~\bibnamefont {D'Elia}}, \bibinfo {author} {\bibfnamefont {M.}~\bibnamefont {Mariti}}, \bibinfo {author} {\bibfnamefont {M.}~\bibnamefont {Mesiti}}, \bibinfo {author} {\bibfnamefont {F.}~\bibnamefont {Negro}},\ and\ \bibinfo {author} {\bibfnamefont {F.}~\bibnamefont {Sanfilippo}},\ }\bibfield  {title} {\bibinfo {title} {{Anisotropy of the quark-antiquark potential in a magnetic field}},\ }\href {https://doi.org/10.1103/PhysRevD.89.114502} {\bibfield  {journal} {\bibinfo  {journal} {Phys. Rev. D}\ }\textbf {\bibinfo {volume} {89}},\ \bibinfo {pages} {114502} (\bibinfo {year} {2014})},\ \Eprint {https://arxiv.org/abs/1403.6094} {arXiv:1403.6094 [hep-lat]} \BibitemShut {NoStop}%
\bibitem [{\citenamefont {Bonati}\ \emph {et~al.}(2016)\citenamefont {Bonati}, \citenamefont {D'Elia}, \citenamefont {Mariti}, \citenamefont {Mesiti}, \citenamefont {Negro}, \citenamefont {Rucci},\ and\ \citenamefont {Sanfilippo}}]{Bonati:2016kxj}%
  \BibitemOpen
  \bibfield  {author} {\bibinfo {author} {\bibfnamefont {C.}~\bibnamefont {Bonati}}, \bibinfo {author} {\bibfnamefont {M.}~\bibnamefont {D'Elia}}, \bibinfo {author} {\bibfnamefont {M.}~\bibnamefont {Mariti}}, \bibinfo {author} {\bibfnamefont {M.}~\bibnamefont {Mesiti}}, \bibinfo {author} {\bibfnamefont {F.}~\bibnamefont {Negro}}, \bibinfo {author} {\bibfnamefont {A.}~\bibnamefont {Rucci}},\ and\ \bibinfo {author} {\bibfnamefont {F.}~\bibnamefont {Sanfilippo}},\ }\bibfield  {title} {\bibinfo {title} {{Magnetic field effects on the static quark potential at zero and finite temperature}},\ }\href {https://doi.org/10.1103/PhysRevD.94.094007} {\bibfield  {journal} {\bibinfo  {journal} {Phys. Rev. D}\ }\textbf {\bibinfo {volume} {94}},\ \bibinfo {pages} {094007} (\bibinfo {year} {2016})},\ \Eprint {https://arxiv.org/abs/1607.08160} {arXiv:1607.08160 [hep-lat]} \BibitemShut {NoStop}%
\bibitem [{\citenamefont {D'Elia}\ \emph {et~al.}(2021)\citenamefont {D'Elia}, \citenamefont {Maio}, \citenamefont {Sanfilippo},\ and\ \citenamefont {Stanzione}}]{DElia:2021tfb}%
  \BibitemOpen
  \bibfield  {author} {\bibinfo {author} {\bibfnamefont {M.}~\bibnamefont {D'Elia}}, \bibinfo {author} {\bibfnamefont {L.}~\bibnamefont {Maio}}, \bibinfo {author} {\bibfnamefont {F.}~\bibnamefont {Sanfilippo}},\ and\ \bibinfo {author} {\bibfnamefont {A.}~\bibnamefont {Stanzione}},\ }\bibfield  {title} {\bibinfo {title} {{Confining and chiral properties of QCD in extremely strong magnetic fields}},\ }\href {https://doi.org/10.1103/PhysRevD.104.114512} {\bibfield  {journal} {\bibinfo  {journal} {Phys. Rev. D}\ }\textbf {\bibinfo {volume} {104}},\ \bibinfo {pages} {114512} (\bibinfo {year} {2021})},\ \Eprint {https://arxiv.org/abs/2109.07456} {arXiv:2109.07456 [hep-lat]} \BibitemShut {NoStop}%
\bibitem [{\citenamefont {Bali}\ \emph {et~al.}(2012)\citenamefont {Bali}, \citenamefont {Bruckmann}, \citenamefont {Endr\ifmmode~\mbox{\H{o}}\else \H{o}\fi{}di}, \citenamefont {Fodor}, \citenamefont {Katz}, \citenamefont {Krieg}, \citenamefont {Schafer},\ and\ \citenamefont {Szabo}}]{Bali:2011qj}%
  \BibitemOpen
  \bibfield  {author} {\bibinfo {author} {\bibfnamefont {G.~S.}\ \bibnamefont {Bali}}, \bibinfo {author} {\bibfnamefont {F.}~\bibnamefont {Bruckmann}}, \bibinfo {author} {\bibfnamefont {G.}~\bibnamefont {Endr\ifmmode~\mbox{\H{o}}\else \H{o}\fi{}di}}, \bibinfo {author} {\bibfnamefont {Z.}~\bibnamefont {Fodor}}, \bibinfo {author} {\bibfnamefont {S.~D.}\ \bibnamefont {Katz}}, \bibinfo {author} {\bibfnamefont {S.}~\bibnamefont {Krieg}}, \bibinfo {author} {\bibfnamefont {A.}~\bibnamefont {Schafer}},\ and\ \bibinfo {author} {\bibfnamefont {K.~K.}\ \bibnamefont {Szabo}},\ }\bibfield  {title} {\bibinfo {title} {{The QCD phase diagram for external magnetic fields}},\ }\href {https://doi.org/10.1007/JHEP02(2012)044} {\bibfield  {journal} {\bibinfo  {journal} {JHEP}\ }\textbf {\bibinfo {volume} {02}},\ \bibinfo {pages} {(2012) 044}},\ \Eprint {https://arxiv.org/abs/1111.4956} {arXiv:1111.4956 [hep-lat]} \BibitemShut {NoStop}%
\bibitem [{\citenamefont {Luschevskaya}\ and\ \citenamefont {Larina}(2014)}]{Luschevskaya:2012xd}%
  \BibitemOpen
  \bibfield  {author} {\bibinfo {author} {\bibfnamefont {E.~V.}\ \bibnamefont {Luschevskaya}}\ and\ \bibinfo {author} {\bibfnamefont {O.~V.}\ \bibnamefont {Larina}},\ }\bibfield  {title} {\bibinfo {title} {{The $\rho$ and $A$ mesons in a strong abelian magnetic field in $SU(2)$ lattice gauge theory}},\ }\href {https://doi.org/10.1016/j.nuclphysb.2014.04.003} {\bibfield  {journal} {\bibinfo  {journal} {Nucl. Phys. B}\ }\textbf {\bibinfo {volume} {884}},\ \bibinfo {pages} {1} (\bibinfo {year} {2014})},\ \Eprint {https://arxiv.org/abs/1203.5699} {arXiv:1203.5699 [hep-lat]} \BibitemShut {NoStop}%
\bibitem [{\citenamefont {Hidaka}\ and\ \citenamefont {Yamamoto}(2013)}]{Hidaka:2012mz}%
  \BibitemOpen
  \bibfield  {author} {\bibinfo {author} {\bibfnamefont {Y.}~\bibnamefont {Hidaka}}\ and\ \bibinfo {author} {\bibfnamefont {A.}~\bibnamefont {Yamamoto}},\ }\bibfield  {title} {\bibinfo {title} {{Charged vector mesons in a strong magnetic field}},\ }\href {https://doi.org/10.1103/PhysRevD.87.094502} {\bibfield  {journal} {\bibinfo  {journal} {Phys. Rev. D}\ }\textbf {\bibinfo {volume} {87}},\ \bibinfo {pages} {094502} (\bibinfo {year} {2013})},\ \Eprint {https://arxiv.org/abs/1209.0007} {arXiv:1209.0007 [hep-ph]} \BibitemShut {NoStop}%
\bibitem [{\citenamefont {Luschevskaya}\ \emph {et~al.}(2015)\citenamefont {Luschevskaya}, \citenamefont {Solovjeva}, \citenamefont {Kochetkov},\ and\ \citenamefont {Teryaev}}]{Luschevskaya:2014lga}%
  \BibitemOpen
  \bibfield  {author} {\bibinfo {author} {\bibfnamefont {E.~V.}\ \bibnamefont {Luschevskaya}}, \bibinfo {author} {\bibfnamefont {O.~E.}\ \bibnamefont {Solovjeva}}, \bibinfo {author} {\bibfnamefont {O.~A.}\ \bibnamefont {Kochetkov}},\ and\ \bibinfo {author} {\bibfnamefont {O.~V.}\ \bibnamefont {Teryaev}},\ }\bibfield  {title} {\bibinfo {title} {{Magnetic polarizabilities of light mesons in $SU(3)$ lattice gauge theory}},\ }\href {https://doi.org/10.1016/j.nuclphysb.2015.07.023} {\bibfield  {journal} {\bibinfo  {journal} {Nucl. Phys. B}\ }\textbf {\bibinfo {volume} {898}},\ \bibinfo {pages} {627} (\bibinfo {year} {2015})},\ \Eprint {https://arxiv.org/abs/1411.4284} {arXiv:1411.4284 [hep-lat]} \BibitemShut {NoStop}%
\bibitem [{\citenamefont {Luschevskaya}\ \emph {et~al.}(2016)\citenamefont {Luschevskaya}, \citenamefont {Solovjeva},\ and\ \citenamefont {Teryaev}}]{Luschevskaya:2015cko}%
  \BibitemOpen
  \bibfield  {author} {\bibinfo {author} {\bibfnamefont {E.~V.}\ \bibnamefont {Luschevskaya}}, \bibinfo {author} {\bibfnamefont {O.~E.}\ \bibnamefont {Solovjeva}},\ and\ \bibinfo {author} {\bibfnamefont {O.~V.}\ \bibnamefont {Teryaev}},\ }\bibfield  {title} {\bibinfo {title} {{Magnetic polarizability of pion}},\ }\href {https://doi.org/10.1016/j.physletb.2016.08.054} {\bibfield  {journal} {\bibinfo  {journal} {Phys. Lett. B}\ }\textbf {\bibinfo {volume} {761}},\ \bibinfo {pages} {393} (\bibinfo {year} {2016})},\ \Eprint {https://arxiv.org/abs/1511.09316} {arXiv:1511.09316 [hep-lat]} \BibitemShut {NoStop}%
\bibitem [{\citenamefont {Bali}\ \emph {et~al.}(2018{\natexlab{a}})\citenamefont {Bali}, \citenamefont {Brandt}, \citenamefont {Endr{\H{o}}di},\ and\ \citenamefont {Gl{\"a}{\ss}le}}]{Bali:2017ian}%
  \BibitemOpen
  \bibfield  {author} {\bibinfo {author} {\bibfnamefont {G.~S.}\ \bibnamefont {Bali}}, \bibinfo {author} {\bibfnamefont {B.~B.}\ \bibnamefont {Brandt}}, \bibinfo {author} {\bibfnamefont {G.}~\bibnamefont {Endr{\H{o}}di}},\ and\ \bibinfo {author} {\bibfnamefont {B.}~\bibnamefont {Gl{\"a}{\ss}le}},\ }\bibfield  {title} {\bibinfo {title} {{Meson masses in electromagnetic fields with Wilson fermions}},\ }\href {https://doi.org/10.1103/PhysRevD.97.034505} {\bibfield  {journal} {\bibinfo  {journal} {Phys. Rev. D}\ }\textbf {\bibinfo {volume} {97}},\ \bibinfo {pages} {034505} (\bibinfo {year} {2018}{\natexlab{a}})},\ \Eprint {https://arxiv.org/abs/1707.05600} {arXiv:1707.05600 [hep-lat]} \BibitemShut {NoStop}%
\bibitem [{\citenamefont {Bali}\ \emph {et~al.}(2018{\natexlab{b}})\citenamefont {Bali}, \citenamefont {Brandt}, \citenamefont {Endr{\H{o}}di},\ and\ \citenamefont {Gl{\"a}{\ss}le}}]{Bali:2018sey}%
  \BibitemOpen
  \bibfield  {author} {\bibinfo {author} {\bibfnamefont {G.~S.}\ \bibnamefont {Bali}}, \bibinfo {author} {\bibfnamefont {B.~B.}\ \bibnamefont {Brandt}}, \bibinfo {author} {\bibfnamefont {G.}~\bibnamefont {Endr{\H{o}}di}},\ and\ \bibinfo {author} {\bibfnamefont {B.}~\bibnamefont {Gl{\"a}{\ss}le}},\ }\bibfield  {title} {\bibinfo {title} {{Weak decay of magnetized pions}},\ }\href {https://doi.org/10.1103/PhysRevLett.121.072001} {\bibfield  {journal} {\bibinfo  {journal} {Phys. Rev. Lett.}\ }\textbf {\bibinfo {volume} {121}},\ \bibinfo {pages} {072001} (\bibinfo {year} {2018}{\natexlab{b}})},\ \Eprint {https://arxiv.org/abs/1805.10971} {arXiv:1805.10971 [hep-lat]} \BibitemShut {NoStop}%
\bibitem [{\citenamefont {Luschevskaya}\ \emph {et~al.}(2018)\citenamefont {Luschevskaya}, \citenamefont {Teryaev}, \citenamefont {Golubkov}, \citenamefont {Solovjeva},\ and\ \citenamefont {Ishkuvatov}}]{Luschevskaya:2018chr}%
  \BibitemOpen
  \bibfield  {author} {\bibinfo {author} {\bibfnamefont {E.~V.}\ \bibnamefont {Luschevskaya}}, \bibinfo {author} {\bibfnamefont {O.~V.}\ \bibnamefont {Teryaev}}, \bibinfo {author} {\bibfnamefont {D.~Y.}\ \bibnamefont {Golubkov}}, \bibinfo {author} {\bibfnamefont {O.~V.}\ \bibnamefont {Solovjeva}},\ and\ \bibinfo {author} {\bibfnamefont {R.~A.}\ \bibnamefont {Ishkuvatov}},\ }\bibfield  {title} {\bibinfo {title} {{Tensor polarizability of the vector mesons from $SU(3)$ lattice gauge theory}},\ }\href {https://doi.org/10.1007/JHEP11(2018)186} {\bibfield  {journal} {\bibinfo  {journal} {JHEP}\ }\textbf {\bibinfo {volume} {11}},\ \bibinfo {pages} {(2018) 186}},\ \Eprint {https://arxiv.org/abs/1811.02344} {arXiv:1811.02344 [hep-lat]} \BibitemShut {NoStop}%
\bibitem [{\citenamefont {Hattori}\ and\ \citenamefont {Yamamoto}(2019)}]{Hattori:2019ijy}%
  \BibitemOpen
  \bibfield  {author} {\bibinfo {author} {\bibfnamefont {K.}~\bibnamefont {Hattori}}\ and\ \bibinfo {author} {\bibfnamefont {A.}~\bibnamefont {Yamamoto}},\ }\bibfield  {title} {\bibinfo {title} {{Meson deformation by magnetic fields in lattice QCD}},\ }\href {https://doi.org/10.1093/ptep/ptz023} {\bibfield  {journal} {\bibinfo  {journal} {PTEP}\ }\textbf {\bibinfo {volume} {2019}},\ \bibinfo {pages} {043B04} (\bibinfo {year} {2019})},\ \Eprint {https://arxiv.org/abs/1901.10182} {arXiv:1901.10182 [hep-lat]} \BibitemShut {NoStop}%
\bibitem [{\citenamefont {Bignell}\ \emph {et~al.}(2019)\citenamefont {Bignell}, \citenamefont {Kamleh},\ and\ \citenamefont {Leinweber}}]{Bignell:2019vpy}%
  \BibitemOpen
  \bibfield  {author} {\bibinfo {author} {\bibfnamefont {R.}~\bibnamefont {Bignell}}, \bibinfo {author} {\bibfnamefont {W.}~\bibnamefont {Kamleh}},\ and\ \bibinfo {author} {\bibfnamefont {D.}~\bibnamefont {Leinweber}},\ }\bibfield  {title} {\bibinfo {title} {{Pion in a uniform background magnetic field with clover fermions}},\ }\href {https://doi.org/10.1103/PhysRevD.100.114518} {\bibfield  {journal} {\bibinfo  {journal} {Phys. Rev. D}\ }\textbf {\bibinfo {volume} {100}},\ \bibinfo {pages} {114518} (\bibinfo {year} {2019})},\ \Eprint {https://arxiv.org/abs/1910.14244} {arXiv:1910.14244 [hep-lat]} \BibitemShut {NoStop}%
\bibitem [{\citenamefont {Bignell}\ \emph {et~al.}(2020)\citenamefont {Bignell}, \citenamefont {Kamleh},\ and\ \citenamefont {Leinweber}}]{Bignell:2020dze}%
  \BibitemOpen
  \bibfield  {author} {\bibinfo {author} {\bibfnamefont {R.}~\bibnamefont {Bignell}}, \bibinfo {author} {\bibfnamefont {W.}~\bibnamefont {Kamleh}},\ and\ \bibinfo {author} {\bibfnamefont {D.}~\bibnamefont {Leinweber}},\ }\bibfield  {title} {\bibinfo {title} {{Pion magnetic polarisability using the background field method}},\ }\href {https://doi.org/10.1016/j.physletb.2020.135853} {\bibfield  {journal} {\bibinfo  {journal} {Phys. Lett. B}\ }\textbf {\bibinfo {volume} {811}},\ \bibinfo {pages} {135853} (\bibinfo {year} {2020})},\ \Eprint {https://arxiv.org/abs/2005.10453} {arXiv:2005.10453 [hep-lat]} \BibitemShut {NoStop}%
\bibitem [{\citenamefont {Ding}\ \emph {et~al.}(2021)\citenamefont {Ding}, \citenamefont {Li}, \citenamefont {Tomiya}, \citenamefont {Wang},\ and\ \citenamefont {Zhang}}]{Ding:2020hxw}%
  \BibitemOpen
  \bibfield  {author} {\bibinfo {author} {\bibfnamefont {H.-T.}\ \bibnamefont {Ding}}, \bibinfo {author} {\bibfnamefont {S.-T.}\ \bibnamefont {Li}}, \bibinfo {author} {\bibfnamefont {A.}~\bibnamefont {Tomiya}}, \bibinfo {author} {\bibfnamefont {X.-D.}\ \bibnamefont {Wang}},\ and\ \bibinfo {author} {\bibfnamefont {Y.}~\bibnamefont {Zhang}},\ }\bibfield  {title} {\bibinfo {title} {{Chiral properties of (2+1)-flavor QCD in strong magnetic fields at zero temperature}},\ }\href {https://doi.org/10.1103/PhysRevD.104.014505} {\bibfield  {journal} {\bibinfo  {journal} {Phys. Rev. D}\ }\textbf {\bibinfo {volume} {104}},\ \bibinfo {pages} {014505} (\bibinfo {year} {2021})},\ \Eprint {https://arxiv.org/abs/2008.00493} {arXiv:2008.00493 [hep-lat]} \BibitemShut {NoStop}%
\bibitem [{\citenamefont {Ding}\ \emph {et~al.}(2022)\citenamefont {Ding}, \citenamefont {Li}, \citenamefont {Liu},\ and\ \citenamefont {Wang}}]{Ding:2022tqn}%
  \BibitemOpen
  \bibfield  {author} {\bibinfo {author} {\bibfnamefont {H.-T.}\ \bibnamefont {Ding}}, \bibinfo {author} {\bibfnamefont {S.-T.}\ \bibnamefont {Li}}, \bibinfo {author} {\bibfnamefont {J.-H.}\ \bibnamefont {Liu}},\ and\ \bibinfo {author} {\bibfnamefont {X.-D.}\ \bibnamefont {Wang}},\ }\bibfield  {title} {\bibinfo {title} {{Chiral condensates and screening masses of neutral pseudoscalar mesons in thermomagnetic QCD medium}},\ }\href {https://doi.org/10.1103/PhysRevD.105.034514} {\bibfield  {journal} {\bibinfo  {journal} {Phys. Rev. D}\ }\textbf {\bibinfo {volume} {105}},\ \bibinfo {pages} {034514} (\bibinfo {year} {2022})},\ \Eprint {https://arxiv.org/abs/2201.02349} {arXiv:2201.02349 [hep-lat]} \BibitemShut {NoStop}%
\bibitem [{\citenamefont {Endr\ifmmode~\mbox{\H{o}}\else \H{o}\fi{}di}(2025)}]{Endrodi:2024cqn}%
  \BibitemOpen
  \bibfield  {author} {\bibinfo {author} {\bibfnamefont {G.}~\bibnamefont {Endr\ifmmode~\mbox{\H{o}}\else \H{o}\fi{}di}},\ }\bibfield  {title} {\bibinfo {title} {{QCD with background electromagnetic fields on the lattice: A review}},\ }\href {https://doi.org/10.1016/j.ppnp.2024.104153} {\bibfield  {journal} {\bibinfo  {journal} {Prog. Part. Nucl. Phys.}\ }\textbf {\bibinfo {volume} {141}},\ \bibinfo {pages} {104153} (\bibinfo {year} {2025})},\ \Eprint {https://arxiv.org/abs/2406.19780} {arXiv:2406.19780 [hep-lat]} \BibitemShut {NoStop}%
\bibitem [{\citenamefont {Ding}\ \emph {et~al.}(2025)\citenamefont {Ding}, \citenamefont {Gu}, \citenamefont {Li},\ and\ \citenamefont {Thakkar}}]{Ding:2025pbu}%
  \BibitemOpen
  \bibfield  {author} {\bibinfo {author} {\bibfnamefont {H.-T.}\ \bibnamefont {Ding}}, \bibinfo {author} {\bibfnamefont {J.-B.}\ \bibnamefont {Gu}}, \bibinfo {author} {\bibfnamefont {S.-T.}\ \bibnamefont {Li}},\ and\ \bibinfo {author} {\bibfnamefont {R.}~\bibnamefont {Thakkar}},\ }\bibfield  {title} {\bibinfo {title} {{Chiral condensates and screening masses of neutral pseudoscalar mesons from lattice QCD at physical quark masses}},\ }\href {https://doi.org/10.1103/PhysRevD.111.074513} {\bibfield  {journal} {\bibinfo  {journal} {Phys. Rev. D}\ }\textbf {\bibinfo {volume} {111}},\ \bibinfo {pages} {074513} (\bibinfo {year} {2025})},\ \Eprint {https://arxiv.org/abs/2501.11262} {arXiv:2501.11262 [hep-lat]} \BibitemShut {NoStop}%
\bibitem [{\citenamefont {Ding}\ and\ \citenamefont {Zhang}(2026)}]{Ding:2026qzu}%
  \BibitemOpen
  \bibfield  {author} {\bibinfo {author} {\bibfnamefont {H.-T.}\ \bibnamefont {Ding}}\ and\ \bibinfo {author} {\bibfnamefont {D.}~\bibnamefont {Zhang}},\ }\bibfield  {title} {\bibinfo {title} {{Chiral properties of (2+1)-flavor QCD in magnetic fields at zero temperature}},\ }\href {https://doi.org/10.1103/qpxq-xjqc} {\bibfield  {journal} {\bibinfo  {journal} {Phys. Rev. D}\ }\textbf {\bibinfo {volume} {113}},\ \bibinfo {pages} {094503} (\bibinfo {year} {2026})},\ \Eprint {https://arxiv.org/abs/2601.18354} {arXiv:2601.18354 [hep-lat]} \BibitemShut {NoStop}%
\bibitem [{\citenamefont {Kharzeev}\ \emph {et~al.}(2008)\citenamefont {Kharzeev}, \citenamefont {McLerran},\ and\ \citenamefont {Warringa}}]{Kharzeev:2007jp}%
  \BibitemOpen
  \bibfield  {author} {\bibinfo {author} {\bibfnamefont {D.~E.}\ \bibnamefont {Kharzeev}}, \bibinfo {author} {\bibfnamefont {L.~D.}\ \bibnamefont {McLerran}},\ and\ \bibinfo {author} {\bibfnamefont {H.~J.}\ \bibnamefont {Warringa}},\ }\bibfield  {title} {\bibinfo {title} {{The effects of topological charge change in heavy ion collisions: $``$Event by event P and CP violation$"$}},\ }\href {https://doi.org/10.1016/j.nuclphysa.2008.02.298} {\bibfield  {journal} {\bibinfo  {journal} {Nucl. Phys. A}\ }\textbf {\bibinfo {volume} {803}},\ \bibinfo {pages} {227} (\bibinfo {year} {2008})},\ \Eprint {https://arxiv.org/abs/0711.0950} {arXiv:0711.0950 [hep-ph]} \BibitemShut {NoStop}%
\bibitem [{\citenamefont {Skokov}\ \emph {et~al.}(2009)\citenamefont {Skokov}, \citenamefont {Illarionov},\ and\ \citenamefont {Toneev}}]{Skokov:2009qp}%
  \BibitemOpen
  \bibfield  {author} {\bibinfo {author} {\bibfnamefont {V.}~\bibnamefont {Skokov}}, \bibinfo {author} {\bibfnamefont {A.~Y.}\ \bibnamefont {Illarionov}},\ and\ \bibinfo {author} {\bibfnamefont {V.}~\bibnamefont {Toneev}},\ }\bibfield  {title} {\bibinfo {title} {{Estimate of the magnetic field strength in heavy-ion collisions}},\ }\href {https://doi.org/10.1142/S0217751X09047570} {\bibfield  {journal} {\bibinfo  {journal} {Int. J. Mod. Phys. A}\ }\textbf {\bibinfo {volume} {24}},\ \bibinfo {pages} {5925} (\bibinfo {year} {2009})},\ \Eprint {https://arxiv.org/abs/0907.1396} {arXiv:0907.1396 [nucl-th]} \BibitemShut {NoStop}%
\bibitem [{\citenamefont {Voronyuk}\ \emph {et~al.}(2011)\citenamefont {Voronyuk}, \citenamefont {Toneev}, \citenamefont {Cassing}, \citenamefont {Bratkovskaya}, \citenamefont {Konchakovski},\ and\ \citenamefont {Voloshin}}]{Voronyuk:2011jd}%
  \BibitemOpen
  \bibfield  {author} {\bibinfo {author} {\bibfnamefont {V.}~\bibnamefont {Voronyuk}}, \bibinfo {author} {\bibfnamefont {V.~D.}\ \bibnamefont {Toneev}}, \bibinfo {author} {\bibfnamefont {W.}~\bibnamefont {Cassing}}, \bibinfo {author} {\bibfnamefont {E.~L.}\ \bibnamefont {Bratkovskaya}}, \bibinfo {author} {\bibfnamefont {V.~P.}\ \bibnamefont {Konchakovski}},\ and\ \bibinfo {author} {\bibfnamefont {S.~A.}\ \bibnamefont {Voloshin}},\ }\bibfield  {title} {\bibinfo {title} {{Electromagnetic field evolution in relativistic heavy-ion collisions}},\ }\href {https://doi.org/10.1103/PhysRevC.83.054911} {\bibfield  {journal} {\bibinfo  {journal} {Phys. Rev. C}\ }\textbf {\bibinfo {volume} {83}},\ \bibinfo {pages} {054911} (\bibinfo {year} {2011})},\ \Eprint {https://arxiv.org/abs/1103.4239} {arXiv:1103.4239 [nucl-th]} \BibitemShut {NoStop}%
\bibitem [{\citenamefont {Deng}\ and\ \citenamefont {Huang}(2012)}]{Deng:2012pc}%
  \BibitemOpen
  \bibfield  {author} {\bibinfo {author} {\bibfnamefont {W.-T.}\ \bibnamefont {Deng}}\ and\ \bibinfo {author} {\bibfnamefont {X.-G.}\ \bibnamefont {Huang}},\ }\bibfield  {title} {\bibinfo {title} {{Event-by-event generation of electromagnetic fields in heavy-ion collisions}},\ }\href {https://doi.org/10.1103/PhysRevC.85.044907} {\bibfield  {journal} {\bibinfo  {journal} {Phys. Rev. C}\ }\textbf {\bibinfo {volume} {85}},\ \bibinfo {pages} {044907} (\bibinfo {year} {2012})},\ \Eprint {https://arxiv.org/abs/1201.5108} {arXiv:1201.5108 [nucl-th]} \BibitemShut {NoStop}%
\bibitem [{\citenamefont {Eichten}\ \emph {et~al.}(1980)\citenamefont {Eichten}, \citenamefont {Gottfried}, \citenamefont {Kinoshita}, \citenamefont {Lane},\ and\ \citenamefont {Yan}}]{Eichten:1979ms}%
  \BibitemOpen
  \bibfield  {author} {\bibinfo {author} {\bibfnamefont {E.}~\bibnamefont {Eichten}}, \bibinfo {author} {\bibfnamefont {K.}~\bibnamefont {Gottfried}}, \bibinfo {author} {\bibfnamefont {T.}~\bibnamefont {Kinoshita}}, \bibinfo {author} {\bibfnamefont {K.~D.}\ \bibnamefont {Lane}},\ and\ \bibinfo {author} {\bibfnamefont {T.~M.}\ \bibnamefont {Yan}},\ }\bibfield  {title} {\bibinfo {title} {{Charmonium: Comparison with Experiment}},\ }\href {https://doi.org/10.1103/PhysRevD.21.203} {\bibfield  {journal} {\bibinfo  {journal} {Phys. Rev. D}\ }\textbf {\bibinfo {volume} {21}},\ \bibinfo {pages} {203} (\bibinfo {year} {1980})}\BibitemShut {NoStop}%
\bibitem [{\citenamefont {Buchmüller}\ and\ \citenamefont {Tye}(1981)}]{Buchmuller:1980su}%
  \BibitemOpen
  \bibfield  {author} {\bibinfo {author} {\bibfnamefont {W.}~\bibnamefont {Buchmüller}}\ and\ \bibinfo {author} {\bibfnamefont {S.~H.~H.}\ \bibnamefont {Tye}},\ }\bibfield  {title} {\bibinfo {title} {{Quarkonia and Quantum Chromodynamics}},\ }\href {https://doi.org/10.1103/PhysRevD.24.132} {\bibfield  {journal} {\bibinfo  {journal} {Phys. Rev. D}\ }\textbf {\bibinfo {volume} {24}},\ \bibinfo {pages} {132} (\bibinfo {year} {1981})}\BibitemShut {NoStop}%
\bibitem [{\citenamefont {Godfrey}\ and\ \citenamefont {Isgur}(1985)}]{Godfrey:1985xj}%
  \BibitemOpen
  \bibfield  {author} {\bibinfo {author} {\bibfnamefont {S.}~\bibnamefont {Godfrey}}\ and\ \bibinfo {author} {\bibfnamefont {N.}~\bibnamefont {Isgur}},\ }\bibfield  {title} {\bibinfo {title} {Mesons in a relativized quark model with chromodynamics},\ }\href {https://doi.org/10.1103/PhysRevD.32.189} {\bibfield  {journal} {\bibinfo  {journal} {Phys. Rev. D}\ }\textbf {\bibinfo {volume} {32}},\ \bibinfo {pages} {189} (\bibinfo {year} {1985})}\BibitemShut {NoStop}%
\bibitem [{\citenamefont {Marasinghe}\ and\ \citenamefont {Tuchin}(2011)}]{Marasinghe:2011bt}%
  \BibitemOpen
  \bibfield  {author} {\bibinfo {author} {\bibfnamefont {K.}~\bibnamefont {Marasinghe}}\ and\ \bibinfo {author} {\bibfnamefont {K.}~\bibnamefont {Tuchin}},\ }\bibfield  {title} {\bibinfo {title} {{Quarkonium dissociation in quark-gluon plasma via ionization in a magnetic field}},\ }\href {https://doi.org/10.1103/PhysRevC.84.044908} {\bibfield  {journal} {\bibinfo  {journal} {Phys. Rev. C}\ }\textbf {\bibinfo {volume} {84}},\ \bibinfo {pages} {044908} (\bibinfo {year} {2011})},\ \Eprint {https://arxiv.org/abs/1103.1329} {arXiv:1103.1329 [hep-ph]} \BibitemShut {NoStop}%
\bibitem [{\citenamefont {Tuchin}(2011)}]{Tuchin:2011cg}%
  \BibitemOpen
  \bibfield  {author} {\bibinfo {author} {\bibfnamefont {K.}~\bibnamefont {Tuchin}},\ }\bibfield  {title} {\bibinfo {title} {{J/{\ensuremath{\psi}} dissociation in parity-odd bubbles}},\ }\href {https://doi.org/10.1016/j.physletb.2011.10.047} {\bibfield  {journal} {\bibinfo  {journal} {Phys. Lett. B}\ }\textbf {\bibinfo {volume} {705}},\ \bibinfo {pages} {482} (\bibinfo {year} {2011})},\ \Eprint {https://arxiv.org/abs/1105.5360} {arXiv:1105.5360 [nucl-th]} \BibitemShut {NoStop}%
\bibitem [{\citenamefont {Yang}\ and\ \citenamefont {M{\"u}ller}(2012)}]{Yang:2011cz}%
  \BibitemOpen
  \bibfield  {author} {\bibinfo {author} {\bibfnamefont {D.-L.}\ \bibnamefont {Yang}}\ and\ \bibinfo {author} {\bibfnamefont {B.}~\bibnamefont {M{\"u}ller}},\ }\bibfield  {title} {\bibinfo {title} {{$J/\psi$ production by magnetic excitation of $\eta_c$}},\ }\href {https://doi.org/10.1088/0954-3899/39/1/015007} {\bibfield  {journal} {\bibinfo  {journal} {J. Phys. G}\ }\textbf {\bibinfo {volume} {39}},\ \bibinfo {pages} {015007} (\bibinfo {year} {2012})},\ \Eprint {https://arxiv.org/abs/1108.2525} {arXiv:1108.2525 [hep-ph]} \BibitemShut {NoStop}%
\bibitem [{\citenamefont {Tuchin}(2013)}]{Tuchin:2013ie}%
  \BibitemOpen
  \bibfield  {author} {\bibinfo {author} {\bibfnamefont {K.}~\bibnamefont {Tuchin}},\ }\bibfield  {title} {\bibinfo {title} {{Particle production in strong electromagnetic fields in relativistic heavy-ion collisions}},\ }\href {https://doi.org/10.1155/2013/490495} {\bibfield  {journal} {\bibinfo  {journal} {Adv. High Energy Phys.}\ }\textbf {\bibinfo {volume} {2013}},\ \bibinfo {pages} {490495} (\bibinfo {year} {2013})},\ \Eprint {https://arxiv.org/abs/1301.0099} {arXiv:1301.0099 [hep-ph]} \BibitemShut {NoStop}%
\bibitem [{\citenamefont {Machado}\ \emph {et~al.}(2013)\citenamefont {Machado}, \citenamefont {Navarra}, \citenamefont {de~Oliveira}, \citenamefont {Noronha},\ and\ \citenamefont {Strickland}}]{Machado:2013rta}%
  \BibitemOpen
  \bibfield  {author} {\bibinfo {author} {\bibfnamefont {C.~S.}\ \bibnamefont {Machado}}, \bibinfo {author} {\bibfnamefont {F.~S.}\ \bibnamefont {Navarra}}, \bibinfo {author} {\bibfnamefont {E.~G.}\ \bibnamefont {de~Oliveira}}, \bibinfo {author} {\bibfnamefont {J.}~\bibnamefont {Noronha}},\ and\ \bibinfo {author} {\bibfnamefont {M.}~\bibnamefont {Strickland}},\ }\bibfield  {title} {\bibinfo {title} {{Heavy quarkonium production in a strong magnetic field}},\ }\href {https://doi.org/10.1103/PhysRevD.88.034009} {\bibfield  {journal} {\bibinfo  {journal} {Phys. Rev. D}\ }\textbf {\bibinfo {volume} {88}},\ \bibinfo {pages} {034009} (\bibinfo {year} {2013})},\ \Eprint {https://arxiv.org/abs/1305.3308} {arXiv:1305.3308 [hep-ph]} \BibitemShut {NoStop}%
\bibitem [{\citenamefont {Alford}\ and\ \citenamefont {Strickland}(2013)}]{Alford:2013jva}%
  \BibitemOpen
  \bibfield  {author} {\bibinfo {author} {\bibfnamefont {J.}~\bibnamefont {Alford}}\ and\ \bibinfo {author} {\bibfnamefont {M.}~\bibnamefont {Strickland}},\ }\bibfield  {title} {\bibinfo {title} {{Charmonia and bottomonia in a magnetic field}},\ }\href {https://doi.org/10.1103/PhysRevD.88.105017} {\bibfield  {journal} {\bibinfo  {journal} {Phys. Rev. D}\ }\textbf {\bibinfo {volume} {88}},\ \bibinfo {pages} {105017} (\bibinfo {year} {2013})},\ \Eprint {https://arxiv.org/abs/1309.3003} {arXiv:1309.3003 [hep-ph]} \BibitemShut {NoStop}%
\bibitem [{\citenamefont {Cho}\ \emph {et~al.}(2014)\citenamefont {Cho}, \citenamefont {Hattori}, \citenamefont {Lee}, \citenamefont {Morita},\ and\ \citenamefont {Ozaki}}]{Cho:2014exa}%
  \BibitemOpen
  \bibfield  {author} {\bibinfo {author} {\bibfnamefont {S.}~\bibnamefont {Cho}}, \bibinfo {author} {\bibfnamefont {K.}~\bibnamefont {Hattori}}, \bibinfo {author} {\bibfnamefont {S.~H.}\ \bibnamefont {Lee}}, \bibinfo {author} {\bibfnamefont {K.}~\bibnamefont {Morita}},\ and\ \bibinfo {author} {\bibfnamefont {S.}~\bibnamefont {Ozaki}},\ }\bibfield  {title} {\bibinfo {title} {{QCD sum rules for magnetically induced mixing between $\eta_c$ and $J/\psi$}},\ }\href {https://doi.org/10.1103/PhysRevLett.113.172301} {\bibfield  {journal} {\bibinfo  {journal} {Phys. Rev. Lett.}\ }\textbf {\bibinfo {volume} {113}},\ \bibinfo {pages} {172301} (\bibinfo {year} {2014})},\ \Eprint {https://arxiv.org/abs/1406.4586} {arXiv:1406.4586 [hep-ph]} \BibitemShut {NoStop}%
\bibitem [{\citenamefont {Dudal}\ and\ \citenamefont {Mertens}(2015)}]{Dudal:2014jfa}%
  \BibitemOpen
  \bibfield  {author} {\bibinfo {author} {\bibfnamefont {D.}~\bibnamefont {Dudal}}\ and\ \bibinfo {author} {\bibfnamefont {T.~G.}\ \bibnamefont {Mertens}},\ }\bibfield  {title} {\bibinfo {title} {{Melting of charmonium in a magnetic field from an effective AdS/QCD model}},\ }\href {https://doi.org/10.1103/PhysRevD.91.086002} {\bibfield  {journal} {\bibinfo  {journal} {Phys. Rev. D}\ }\textbf {\bibinfo {volume} {91}},\ \bibinfo {pages} {086002} (\bibinfo {year} {2015})},\ \Eprint {https://arxiv.org/abs/1410.3297} {arXiv:1410.3297 [hep-th]} \BibitemShut {NoStop}%
\bibitem [{\citenamefont {Cho}\ \emph {et~al.}(2015)\citenamefont {Cho}, \citenamefont {Hattori}, \citenamefont {Lee}, \citenamefont {Morita},\ and\ \citenamefont {Ozaki}}]{Cho:2014loa}%
  \BibitemOpen
  \bibfield  {author} {\bibinfo {author} {\bibfnamefont {S.}~\bibnamefont {Cho}}, \bibinfo {author} {\bibfnamefont {K.}~\bibnamefont {Hattori}}, \bibinfo {author} {\bibfnamefont {S.~H.}\ \bibnamefont {Lee}}, \bibinfo {author} {\bibfnamefont {K.}~\bibnamefont {Morita}},\ and\ \bibinfo {author} {\bibfnamefont {S.}~\bibnamefont {Ozaki}},\ }\bibfield  {title} {\bibinfo {title} {{Charmonium Spectroscopy in Strong Magnetic Fields by QCD Sum Rules: $S$-Wave Ground States}},\ }\href {https://doi.org/10.1103/PhysRevD.91.045025} {\bibfield  {journal} {\bibinfo  {journal} {Phys. Rev. D}\ }\textbf {\bibinfo {volume} {91}},\ \bibinfo {pages} {045025} (\bibinfo {year} {2015})},\ \Eprint {https://arxiv.org/abs/1411.7675} {arXiv:1411.7675 [hep-ph]} \BibitemShut {NoStop}%
\bibitem [{\citenamefont {Bonati}\ \emph {et~al.}(2015)\citenamefont {Bonati}, \citenamefont {D'Elia},\ and\ \citenamefont {Rucci}}]{Bonati:2015dka}%
  \BibitemOpen
  \bibfield  {author} {\bibinfo {author} {\bibfnamefont {C.}~\bibnamefont {Bonati}}, \bibinfo {author} {\bibfnamefont {M.}~\bibnamefont {D'Elia}},\ and\ \bibinfo {author} {\bibfnamefont {A.}~\bibnamefont {Rucci}},\ }\bibfield  {title} {\bibinfo {title} {{Heavy quarkonia in strong magnetic fields}},\ }\href {https://doi.org/10.1103/PhysRevD.92.054014} {\bibfield  {journal} {\bibinfo  {journal} {Phys. Rev. D}\ }\textbf {\bibinfo {volume} {92}},\ \bibinfo {pages} {054014} (\bibinfo {year} {2015})},\ \Eprint {https://arxiv.org/abs/1506.07890} {arXiv:1506.07890 [hep-ph]} \BibitemShut {NoStop}%
\bibitem [{\citenamefont {Sadofyev}\ and\ \citenamefont {Yin}(2016)}]{Sadofyev:2015hxa}%
  \BibitemOpen
  \bibfield  {author} {\bibinfo {author} {\bibfnamefont {A.~V.}\ \bibnamefont {Sadofyev}}\ and\ \bibinfo {author} {\bibfnamefont {Y.}~\bibnamefont {Yin}},\ }\bibfield  {title} {\bibinfo {title} {{The charmonium dissociation in an {\textquotedblleft}anomalous wind{\textquotedblright}}},\ }\href {https://doi.org/10.1007/JHEP01(2016)052} {\bibfield  {journal} {\bibinfo  {journal} {JHEP}\ }\textbf {\bibinfo {volume} {01}},\ \bibinfo {pages} {(2016) 052}},\ \Eprint {https://arxiv.org/abs/1510.06760} {arXiv:1510.06760 [hep-th]} \BibitemShut {NoStop}%
\bibitem [{\citenamefont {Suzuki}\ and\ \citenamefont {Yoshida}(2016)}]{Suzuki:2016kcs}%
  \BibitemOpen
  \bibfield  {author} {\bibinfo {author} {\bibfnamefont {K.}~\bibnamefont {Suzuki}}\ and\ \bibinfo {author} {\bibfnamefont {T.}~\bibnamefont {Yoshida}},\ }\bibfield  {title} {\bibinfo {title} {{Cigar-shaped quarkonia under strong magnetic field}},\ }\href {https://doi.org/10.1103/PhysRevD.93.051502} {\bibfield  {journal} {\bibinfo  {journal} {Phys. Rev. D}\ }\textbf {\bibinfo {volume} {93}},\ \bibinfo {pages} {051502} (\bibinfo {year} {2016})},\ \Eprint {https://arxiv.org/abs/1601.02178} {arXiv:1601.02178 [hep-ph]} \BibitemShut {NoStop}%
\bibitem [{\citenamefont {Yoshida}\ and\ \citenamefont {Suzuki}(2016)}]{Yoshida:2016xgm}%
  \BibitemOpen
  \bibfield  {author} {\bibinfo {author} {\bibfnamefont {T.}~\bibnamefont {Yoshida}}\ and\ \bibinfo {author} {\bibfnamefont {K.}~\bibnamefont {Suzuki}},\ }\bibfield  {title} {\bibinfo {title} {{Heavy meson spectroscopy under strong magnetic field}},\ }\href {https://doi.org/10.1103/PhysRevD.94.074043} {\bibfield  {journal} {\bibinfo  {journal} {Phys. Rev. D}\ }\textbf {\bibinfo {volume} {94}},\ \bibinfo {pages} {074043} (\bibinfo {year} {2016})},\ \Eprint {https://arxiv.org/abs/1607.04935} {arXiv:1607.04935 [hep-ph]} \BibitemShut {NoStop}%
\bibitem [{\citenamefont {Hasan}\ \emph {et~al.}(2017)\citenamefont {Hasan}, \citenamefont {Chatterjee},\ and\ \citenamefont {Patra}}]{Hasan:2017fmf}%
  \BibitemOpen
  \bibfield  {author} {\bibinfo {author} {\bibfnamefont {M.}~\bibnamefont {Hasan}}, \bibinfo {author} {\bibfnamefont {B.}~\bibnamefont {Chatterjee}},\ and\ \bibinfo {author} {\bibfnamefont {B.~K.}\ \bibnamefont {Patra}},\ }\bibfield  {title} {\bibinfo {title} {{Heavy quark potential in a static and strong homogeneous magnetic field}},\ }\href {https://doi.org/10.1140/epjc/s10052-017-5346-z} {\bibfield  {journal} {\bibinfo  {journal} {Eur. Phys. J. C}\ }\textbf {\bibinfo {volume} {77}},\ \bibinfo {pages} {767} (\bibinfo {year} {2017})},\ \Eprint {https://arxiv.org/abs/1703.10508} {arXiv:1703.10508 [hep-ph]} \BibitemShut {NoStop}%
\bibitem [{\citenamefont {Singh}\ \emph {et~al.}(2018)\citenamefont {Singh}, \citenamefont {Thakur},\ and\ \citenamefont {Mishra}}]{Singh:2017nfa}%
  \BibitemOpen
  \bibfield  {author} {\bibinfo {author} {\bibfnamefont {B.}~\bibnamefont {Singh}}, \bibinfo {author} {\bibfnamefont {L.}~\bibnamefont {Thakur}},\ and\ \bibinfo {author} {\bibfnamefont {H.}~\bibnamefont {Mishra}},\ }\bibfield  {title} {\bibinfo {title} {{Heavy quark complex potential in a strongly magnetized hot QGP medium}},\ }\href {https://doi.org/10.1103/PhysRevD.97.096011} {\bibfield  {journal} {\bibinfo  {journal} {Phys. Rev. D}\ }\textbf {\bibinfo {volume} {97}},\ \bibinfo {pages} {096011} (\bibinfo {year} {2018})},\ \Eprint {https://arxiv.org/abs/1711.03071} {arXiv:1711.03071 [hep-ph]} \BibitemShut {NoStop}%
\bibitem [{\citenamefont {Braga}\ and\ \citenamefont {Ferreira}(2018)}]{Braga:2018zlu}%
  \BibitemOpen
  \bibfield  {author} {\bibinfo {author} {\bibfnamefont {N.~R.~F.}\ \bibnamefont {Braga}}\ and\ \bibinfo {author} {\bibfnamefont {L.~F.}\ \bibnamefont {Ferreira}},\ }\bibfield  {title} {\bibinfo {title} {{Heavy meson dissociation in a plasma with magnetic fields}},\ }\href {https://doi.org/10.1016/j.physletb.2018.06.053} {\bibfield  {journal} {\bibinfo  {journal} {Phys. Lett. B}\ }\textbf {\bibinfo {volume} {783}},\ \bibinfo {pages} {186} (\bibinfo {year} {2018})},\ \Eprint {https://arxiv.org/abs/1802.02084} {arXiv:1802.02084 [hep-ph]} \BibitemShut {NoStop}%
\bibitem [{\citenamefont {Iwasaki}\ \emph {et~al.}(2019)\citenamefont {Iwasaki}, \citenamefont {Oka}, \citenamefont {Suzuki},\ and\ \citenamefont {Yoshida}}]{Iwasaki:2018pby}%
  \BibitemOpen
  \bibfield  {author} {\bibinfo {author} {\bibfnamefont {S.}~\bibnamefont {Iwasaki}}, \bibinfo {author} {\bibfnamefont {M.}~\bibnamefont {Oka}}, \bibinfo {author} {\bibfnamefont {K.}~\bibnamefont {Suzuki}},\ and\ \bibinfo {author} {\bibfnamefont {T.}~\bibnamefont {Yoshida}},\ }\bibfield  {title} {\bibinfo {title} {{Hadronic Paschen{\textendash}Back effect}},\ }\href {https://doi.org/10.1016/j.physletb.2018.10.024} {\bibfield  {journal} {\bibinfo  {journal} {Phys. Lett. B}\ }\textbf {\bibinfo {volume} {790}},\ \bibinfo {pages} {71} (\bibinfo {year} {2019})},\ \Eprint {https://arxiv.org/abs/1802.04971} {arXiv:1802.04971 [hep-ph]} \BibitemShut {NoStop}%
\bibitem [{\citenamefont {Jahan C.~S.}\ \emph {et~al.}(2018)\citenamefont {Jahan C.~S.}, \citenamefont {Dhale}, \citenamefont {Reddy~P.}, \citenamefont {Kesarwani},\ and\ \citenamefont {Mishra}}]{Amal:2018qln}%
  \BibitemOpen
  \bibfield  {author} {\bibinfo {author} {\bibfnamefont {A.}~\bibnamefont {Jahan C.~S.}}, \bibinfo {author} {\bibfnamefont {N.}~\bibnamefont {Dhale}}, \bibinfo {author} {\bibfnamefont {S.}~\bibnamefont {Reddy~P.}}, \bibinfo {author} {\bibfnamefont {S.}~\bibnamefont {Kesarwani}},\ and\ \bibinfo {author} {\bibfnamefont {A.}~\bibnamefont {Mishra}},\ }\bibfield  {title} {\bibinfo {title} {{Charmonium states in strong magnetic fields}},\ }\href {https://doi.org/10.1103/PhysRevC.98.065202} {\bibfield  {journal} {\bibinfo  {journal} {Phys. Rev. C}\ }\textbf {\bibinfo {volume} {98}},\ \bibinfo {pages} {065202} (\bibinfo {year} {2018})},\ \Eprint {https://arxiv.org/abs/1803.04322} {arXiv:1803.04322 [nucl-th]} \BibitemShut {NoStop}%
\bibitem [{\citenamefont {Iwasaki}\ and\ \citenamefont {Suzuki}(2018)}]{Iwasaki:2018czv}%
  \BibitemOpen
  \bibfield  {author} {\bibinfo {author} {\bibfnamefont {S.}~\bibnamefont {Iwasaki}}\ and\ \bibinfo {author} {\bibfnamefont {K.}~\bibnamefont {Suzuki}},\ }\bibfield  {title} {\bibinfo {title} {{Quarkonium radiative decays from the hadronic Paschen-Back effect}},\ }\href {https://doi.org/10.1103/PhysRevD.98.054017} {\bibfield  {journal} {\bibinfo  {journal} {Phys. Rev. D}\ }\textbf {\bibinfo {volume} {98}},\ \bibinfo {pages} {054017} (\bibinfo {year} {2018})},\ \Eprint {https://arxiv.org/abs/1805.09787} {arXiv:1805.09787 [hep-ph]} \BibitemShut {NoStop}%
\bibitem [{\citenamefont {Hasan}\ \emph {et~al.}(2020)\citenamefont {Hasan}, \citenamefont {Patra}, \citenamefont {Chatterjee},\ and\ \citenamefont {Bagchi}}]{Hasan:2018kvx}%
  \BibitemOpen
  \bibfield  {author} {\bibinfo {author} {\bibfnamefont {M.}~\bibnamefont {Hasan}}, \bibinfo {author} {\bibfnamefont {B.~K.}\ \bibnamefont {Patra}}, \bibinfo {author} {\bibfnamefont {B.}~\bibnamefont {Chatterjee}},\ and\ \bibinfo {author} {\bibfnamefont {P.}~\bibnamefont {Bagchi}},\ }\bibfield  {title} {\bibinfo {title} {{Landau Damping in a strong magnetic field: Dissociation of Quarkonia}},\ }\href {https://doi.org/10.1016/j.nuclphysa.2019.121688} {\bibfield  {journal} {\bibinfo  {journal} {Nucl. Phys. A}\ }\textbf {\bibinfo {volume} {995}},\ \bibinfo {pages} {121688} (\bibinfo {year} {2020})},\ \Eprint {https://arxiv.org/abs/1802.06874} {arXiv:1802.06874 [hep-ph]} \BibitemShut {NoStop}%
\bibitem [{\citenamefont {Braga}\ and\ \citenamefont {Ferreira}(2019)}]{Braga:2019yeh}%
  \BibitemOpen
  \bibfield  {author} {\bibinfo {author} {\bibfnamefont {N.~R.~F.}\ \bibnamefont {Braga}}\ and\ \bibinfo {author} {\bibfnamefont {L.~F.}\ \bibnamefont {Ferreira}},\ }\bibfield  {title} {\bibinfo {title} {{Quasinormal modes for quarkonium in a plasma with magnetic fields}},\ }\href {https://doi.org/10.1016/j.physletb.2019.06.050} {\bibfield  {journal} {\bibinfo  {journal} {Phys. Lett. B}\ }\textbf {\bibinfo {volume} {795}},\ \bibinfo {pages} {462} (\bibinfo {year} {2019})},\ \Eprint {https://arxiv.org/abs/1905.11309} {arXiv:1905.11309 [hep-ph]} \BibitemShut {NoStop}%
\bibitem [{\citenamefont {Hasan}\ and\ \citenamefont {Patra}(2020)}]{Hasan:2020iwa}%
  \BibitemOpen
  \bibfield  {author} {\bibinfo {author} {\bibfnamefont {M.}~\bibnamefont {Hasan}}\ and\ \bibinfo {author} {\bibfnamefont {B.~K.}\ \bibnamefont {Patra}},\ }\bibfield  {title} {\bibinfo {title} {{Dissociation of heavy quarkonia in a weak magnetic field}},\ }\href {https://doi.org/10.1103/PhysRevD.102.036020} {\bibfield  {journal} {\bibinfo  {journal} {Phys. Rev. D}\ }\textbf {\bibinfo {volume} {102}},\ \bibinfo {pages} {036020} (\bibinfo {year} {2020})},\ \Eprint {https://arxiv.org/abs/2004.12857} {arXiv:2004.12857 [hep-ph]} \BibitemShut {NoStop}%
\bibitem [{\citenamefont {Zhou}\ \emph {et~al.}(2020)\citenamefont {Zhou}, \citenamefont {Chen}, \citenamefont {Zhao},\ and\ \citenamefont {Ping}}]{Zhou:2020ssi}%
  \BibitemOpen
  \bibfield  {author} {\bibinfo {author} {\bibfnamefont {J.}~\bibnamefont {Zhou}}, \bibinfo {author} {\bibfnamefont {X.}~\bibnamefont {Chen}}, \bibinfo {author} {\bibfnamefont {Y.-Q.}\ \bibnamefont {Zhao}},\ and\ \bibinfo {author} {\bibfnamefont {J.}~\bibnamefont {Ping}},\ }\bibfield  {title} {\bibinfo {title} {{Thermodynamics of heavy quarkonium in a magnetic field background}},\ }\href {https://doi.org/10.1103/PhysRevD.102.086020} {\bibfield  {journal} {\bibinfo  {journal} {Phys. Rev. D}\ }\textbf {\bibinfo {volume} {102}},\ \bibinfo {pages} {086020} (\bibinfo {year} {2020})},\ \Eprint {https://arxiv.org/abs/2006.09062} {arXiv:2006.09062 [hep-ph]} \BibitemShut {NoStop}%
\bibitem [{\citenamefont {Braga}\ and\ \citenamefont {da~Mata}(2020)}]{Braga:2020hhs}%
  \BibitemOpen
  \bibfield  {author} {\bibinfo {author} {\bibfnamefont {N.~R.~F.}\ \bibnamefont {Braga}}\ and\ \bibinfo {author} {\bibfnamefont {R.}~\bibnamefont {da~Mata}},\ }\bibfield  {title} {\bibinfo {title} {{Configuration entropy description of charmonium dissociation under the influence of magnetic fields}},\ }\href {https://doi.org/10.1016/j.physletb.2020.135918} {\bibfield  {journal} {\bibinfo  {journal} {Phys. Lett. B}\ }\textbf {\bibinfo {volume} {811}},\ \bibinfo {pages} {135918} (\bibinfo {year} {2020})},\ \Eprint {https://arxiv.org/abs/2008.10457} {arXiv:2008.10457 [hep-th]} \BibitemShut {NoStop}%
\bibitem [{\citenamefont {Braga}\ \emph {et~al.}(2022)\citenamefont {Braga}, \citenamefont {Ferreira},\ and\ \citenamefont {Ferreira}}]{Braga:2021fey}%
  \BibitemOpen
  \bibfield  {author} {\bibinfo {author} {\bibfnamefont {N.~R.~F.}\ \bibnamefont {Braga}}, \bibinfo {author} {\bibfnamefont {Y.~F.}\ \bibnamefont {Ferreira}},\ and\ \bibinfo {author} {\bibfnamefont {L.~F.}\ \bibnamefont {Ferreira}},\ }\bibfield  {title} {\bibinfo {title} {{Configuration entropy and stability of bottomonium radial excitations in a plasma with magnetic fields}},\ }\href {https://doi.org/10.1103/PhysRevD.105.114044} {\bibfield  {journal} {\bibinfo  {journal} {Phys. Rev. D}\ }\textbf {\bibinfo {volume} {105}},\ \bibinfo {pages} {114044} (\bibinfo {year} {2022})},\ \Eprint {https://arxiv.org/abs/2110.04560} {arXiv:2110.04560 [hep-th]} \BibitemShut {NoStop}%
\bibitem [{\citenamefont {Jena}\ \emph {et~al.}(2022)\citenamefont {Jena}, \citenamefont {Shukla}, \citenamefont {Dudal},\ and\ \citenamefont {Mahapatra}}]{Jena:2022nzw}%
  \BibitemOpen
  \bibfield  {author} {\bibinfo {author} {\bibfnamefont {S.~S.}\ \bibnamefont {Jena}}, \bibinfo {author} {\bibfnamefont {B.}~\bibnamefont {Shukla}}, \bibinfo {author} {\bibfnamefont {D.}~\bibnamefont {Dudal}},\ and\ \bibinfo {author} {\bibfnamefont {S.}~\bibnamefont {Mahapatra}},\ }\bibfield  {title} {\bibinfo {title} {{Entropic force and real-time dynamics of holographic quarkonium in a magnetic field}},\ }\href {https://doi.org/10.1103/PhysRevD.105.086011} {\bibfield  {journal} {\bibinfo  {journal} {Phys. Rev. D}\ }\textbf {\bibinfo {volume} {105}},\ \bibinfo {pages} {086011} (\bibinfo {year} {2022})},\ \Eprint {https://arxiv.org/abs/2202.01486} {arXiv:2202.01486 [hep-th]} \BibitemShut {NoStop}%
\bibitem [{\citenamefont {Hu}\ \emph {et~al.}(2022)\citenamefont {Hu}, \citenamefont {Shi}, \citenamefont {Xu}, \citenamefont {Zhao},\ and\ \citenamefont {Zhuang}}]{Hu:2022ofv}%
  \BibitemOpen
  \bibfield  {author} {\bibinfo {author} {\bibfnamefont {J.}~\bibnamefont {Hu}}, \bibinfo {author} {\bibfnamefont {S.}~\bibnamefont {Shi}}, \bibinfo {author} {\bibfnamefont {Z.}~\bibnamefont {Xu}}, \bibinfo {author} {\bibfnamefont {J.}~\bibnamefont {Zhao}},\ and\ \bibinfo {author} {\bibfnamefont {P.}~\bibnamefont {Zhuang}},\ }\bibfield  {title} {\bibinfo {title} {{Magnetic field induced hair structure in the charmonium gluon dissociation}},\ }\href {https://doi.org/10.1103/PhysRevD.105.094013} {\bibfield  {journal} {\bibinfo  {journal} {Phys. Rev. D}\ }\textbf {\bibinfo {volume} {105}},\ \bibinfo {pages} {094013} (\bibinfo {year} {2022})},\ \Eprint {https://arxiv.org/abs/2202.07938} {arXiv:2202.07938 [nucl-th]} \BibitemShut {NoStop}%
\bibitem [{\citenamefont {Ghosh}\ \emph {et~al.}(2022)\citenamefont {Ghosh}, \citenamefont {Bandyopadhyay}, \citenamefont {Nilima},\ and\ \citenamefont {Ghosh}}]{Ghosh:2022sxi}%
  \BibitemOpen
  \bibfield  {author} {\bibinfo {author} {\bibfnamefont {R.}~\bibnamefont {Ghosh}}, \bibinfo {author} {\bibfnamefont {A.}~\bibnamefont {Bandyopadhyay}}, \bibinfo {author} {\bibfnamefont {I.}~\bibnamefont {Nilima}},\ and\ \bibinfo {author} {\bibfnamefont {S.}~\bibnamefont {Ghosh}},\ }\bibfield  {title} {\bibinfo {title} {{Anisotropic tomography of heavy quark dissociation by using the general propagator structure in a finite magnetic field}},\ }\href {https://doi.org/10.1103/PhysRevD.106.054010} {\bibfield  {journal} {\bibinfo  {journal} {Phys. Rev. D}\ }\textbf {\bibinfo {volume} {106}},\ \bibinfo {pages} {054010} (\bibinfo {year} {2022})},\ \Eprint {https://arxiv.org/abs/2204.02312} {arXiv:2204.02312 [hep-ph]} \BibitemShut {NoStop}%
\bibitem [{\citenamefont {Parui}\ \emph {et~al.}(2022)\citenamefont {Parui}, \citenamefont {De}, \citenamefont {Kumar},\ and\ \citenamefont {Mishra}}]{Parui:2022msu}%
  \BibitemOpen
  \bibfield  {author} {\bibinfo {author} {\bibfnamefont {P.}~\bibnamefont {Parui}}, \bibinfo {author} {\bibfnamefont {S.}~\bibnamefont {De}}, \bibinfo {author} {\bibfnamefont {A.}~\bibnamefont {Kumar}},\ and\ \bibinfo {author} {\bibfnamefont {A.}~\bibnamefont {Mishra}},\ }\bibfield  {title} {\bibinfo {title} {{QCD sum rule analysis of heavy quarkonium states in magnetized matter: Effects of magnetic and inverse magnetic catalysis}},\ }\href {https://doi.org/10.1103/PhysRevD.106.114033} {\bibfield  {journal} {\bibinfo  {journal} {Phys. Rev. D}\ }\textbf {\bibinfo {volume} {106}},\ \bibinfo {pages} {114033} (\bibinfo {year} {2022})},\ \Eprint {https://arxiv.org/abs/2208.05856} {arXiv:2208.05856 [hep-ph]} \BibitemShut {NoStop}%
\bibitem [{\citenamefont {Sebastian}\ \emph {et~al.}(2023)\citenamefont {Sebastian}, \citenamefont {Thakur}, \citenamefont {Mishra},\ and\ \citenamefont {Haque}}]{Sebastian:2023tlw}%
  \BibitemOpen
  \bibfield  {author} {\bibinfo {author} {\bibfnamefont {J.}~\bibnamefont {Sebastian}}, \bibinfo {author} {\bibfnamefont {L.}~\bibnamefont {Thakur}}, \bibinfo {author} {\bibfnamefont {H.}~\bibnamefont {Mishra}},\ and\ \bibinfo {author} {\bibfnamefont {N.}~\bibnamefont {Haque}},\ }\bibfield  {title} {\bibinfo {title} {{Heavy quarkonia in QGP medium in an arbitrary magnetic field}},\ }\href {https://doi.org/10.1103/PhysRevD.108.094001} {\bibfield  {journal} {\bibinfo  {journal} {Phys. Rev. D}\ }\textbf {\bibinfo {volume} {108}},\ \bibinfo {pages} {094001} (\bibinfo {year} {2023})},\ \Eprint {https://arxiv.org/abs/2308.04410} {arXiv:2308.04410 [hep-ph]} \BibitemShut {NoStop}%
\bibitem [{\citenamefont {Nilima}\ \emph {et~al.}(2024)\citenamefont {Nilima}, \citenamefont {Hasan}, \citenamefont {Singh},\ and\ \citenamefont {Jamal}}]{Nilima:2024nvd}%
  \BibitemOpen
  \bibfield  {author} {\bibinfo {author} {\bibfnamefont {I.}~\bibnamefont {Nilima}}, \bibinfo {author} {\bibfnamefont {M.}~\bibnamefont {Hasan}}, \bibinfo {author} {\bibfnamefont {B.~K.}\ \bibnamefont {Singh}},\ and\ \bibinfo {author} {\bibfnamefont {M.~Y.}\ \bibnamefont {Jamal}},\ }\bibfield  {title} {\bibinfo {title} {{Quarkonia dissociation at finite magnetic field in the presence of momentum anisotropy}},\ }\href {https://doi.org/10.1140/epjc/s10052-024-12525-y} {\bibfield  {journal} {\bibinfo  {journal} {Eur. Phys. J. C}\ }\textbf {\bibinfo {volume} {84}},\ \bibinfo {pages} {160} (\bibinfo {year} {2024})},\ \Eprint {https://arxiv.org/abs/2402.07848} {arXiv:2402.07848 [hep-ph]} \BibitemShut {NoStop}%
\bibitem [{\citenamefont {Jena}\ \emph {et~al.}(2024)\citenamefont {Jena}, \citenamefont {Barman}, \citenamefont {Toniato}, \citenamefont {Dudal},\ and\ \citenamefont {Mahapatra}}]{Jena:2024cqs}%
  \BibitemOpen
  \bibfield  {author} {\bibinfo {author} {\bibfnamefont {S.~S.}\ \bibnamefont {Jena}}, \bibinfo {author} {\bibfnamefont {J.}~\bibnamefont {Barman}}, \bibinfo {author} {\bibfnamefont {B.}~\bibnamefont {Toniato}}, \bibinfo {author} {\bibfnamefont {D.}~\bibnamefont {Dudal}},\ and\ \bibinfo {author} {\bibfnamefont {S.}~\bibnamefont {Mahapatra}},\ }\bibfield  {title} {\bibinfo {title} {{A dynamical Einstein-Born-Infeld-dilaton model and holographic quarkonium melting in a magnetic field}},\ }\href {https://doi.org/10.1007/JHEP12(2024)096} {\bibfield  {journal} {\bibinfo  {journal} {JHEP}\ }\textbf {\bibinfo {volume} {12}},\ \bibinfo {pages} {(2024) 096}},\ \Eprint {https://arxiv.org/abs/2408.14813} {arXiv:2408.14813 [hep-th]} \BibitemShut {NoStop}%
\bibitem [{\citenamefont {Shukla}\ \emph {et~al.}(2025)\citenamefont {Shukla}, \citenamefont {Nongmaithem}, \citenamefont {Dudal},\ and\ \citenamefont {Mahapatra}}]{Shukla:2024qlf}%
  \BibitemOpen
  \bibfield  {author} {\bibinfo {author} {\bibfnamefont {B.}~\bibnamefont {Shukla}}, \bibinfo {author} {\bibfnamefont {J.}~\bibnamefont {Nongmaithem}}, \bibinfo {author} {\bibfnamefont {D.}~\bibnamefont {Dudal}},\ and\ \bibinfo {author} {\bibfnamefont {S.}~\bibnamefont {Mahapatra}},\ }\bibfield  {title} {\bibinfo {title} {{Interplay of magnetic field and chemical potential induced anisotropy and frame dependent chaos of a $Q\bar{Q}$ pair in holographic QCD}},\ }\href {https://doi.org/10.1103/PhysRevD.111.106002} {\bibfield  {journal} {\bibinfo  {journal} {Phys. Rev. D}\ }\textbf {\bibinfo {volume} {111}},\ \bibinfo {pages} {106002} (\bibinfo {year} {2025})},\ \Eprint {https://arxiv.org/abs/2411.17279} {arXiv:2411.17279 [hep-th]} \BibitemShut {NoStop}%
\bibitem [{\citenamefont {Wen}\ \emph {et~al.}(2025)\citenamefont {Wen}, \citenamefont {Li}, \citenamefont {Zhou}, \citenamefont {Li},\ and\ \citenamefont {Vary}}]{Wen:2025dwy}%
  \BibitemOpen
  \bibfield  {author} {\bibinfo {author} {\bibfnamefont {L.}~\bibnamefont {Wen}}, \bibinfo {author} {\bibfnamefont {M.}~\bibnamefont {Li}}, \bibinfo {author} {\bibfnamefont {Y.}~\bibnamefont {Zhou}}, \bibinfo {author} {\bibfnamefont {Y.}~\bibnamefont {Li}},\ and\ \bibinfo {author} {\bibfnamefont {J.~P.}\ \bibnamefont {Vary}},\ }\bibfield  {title} {\bibinfo {title} {{Relativistic dynamics of charmonia in strong magnetic fields}},\ }\href {https://doi.org/10.1103/dnq8-ncd7} {\bibfield  {journal} {\bibinfo  {journal} {Phys. Rev. D}\ }\textbf {\bibinfo {volume} {112}},\ \bibinfo {pages} {014019} (\bibinfo {year} {2025})},\ \Eprint {https://arxiv.org/abs/2504.03294} {arXiv:2504.03294 [hep-ph]} \BibitemShut {NoStop}%
\bibitem [{\citenamefont {Jena}\ \emph {et~al.}(2025)\citenamefont {Jena}, \citenamefont {Bhattacharjee}, \citenamefont {Dudal},\ and\ \citenamefont {Mahapatra}}]{Jena:2025xcf}%
  \BibitemOpen
  \bibfield  {author} {\bibinfo {author} {\bibfnamefont {S.~S.}\ \bibnamefont {Jena}}, \bibinfo {author} {\bibfnamefont {A.}~\bibnamefont {Bhattacharjee}}, \bibinfo {author} {\bibfnamefont {D.}~\bibnamefont {Dudal}},\ and\ \bibinfo {author} {\bibfnamefont {S.}~\bibnamefont {Mahapatra}},\ }\bibfield  {title} {\bibinfo {title} {{Probing quarkonium diffusion in a magnetized quark-gluon plasma}},\ }\href {https://doi.org/10.1103/15z1-bxlm} {\bibfield  {journal} {\bibinfo  {journal} {Phys. Rev. D}\ }\textbf {\bibinfo {volume} {112}},\ \bibinfo {pages} {086010} (\bibinfo {year} {2025})},\ \Eprint {https://arxiv.org/abs/2507.00746} {arXiv:2507.00746 [hep-th]} \BibitemShut {NoStop}%
\bibitem [{\citenamefont {Arifi}\ and\ \citenamefont {Suzuki}(2025)}]{Arifi:2025ivt}%
  \BibitemOpen
  \bibfield  {author} {\bibinfo {author} {\bibfnamefont {A.~J.}\ \bibnamefont {Arifi}}\ and\ \bibinfo {author} {\bibfnamefont {K.}~\bibnamefont {Suzuki}},\ }\bibfield  {title} {\bibinfo {title} {{Structure of heavy quarkonia in a strong magnetic field}},\ }\href {https://doi.org/10.1103/pwsl-xrq5} {\bibfield  {journal} {\bibinfo  {journal} {Phys. Rev. D}\ }\textbf {\bibinfo {volume} {112}},\ \bibinfo {pages} {094013} (\bibinfo {year} {2025})},\ \Eprint {https://arxiv.org/abs/2507.18894} {arXiv:2507.18894 [hep-ph]} \BibitemShut {NoStop}%
\bibitem [{\citenamefont {Dominguez}\ \emph {et~al.}(2025)\citenamefont {Dominguez}, \citenamefont {Koning},\ and\ \citenamefont {Hern{\'a}ndez}}]{Dominguez:2025nar}%
  \BibitemOpen
  \bibfield  {author} {\bibinfo {author} {\bibfnamefont {C.~A.}\ \bibnamefont {Dominguez}}, \bibinfo {author} {\bibfnamefont {M.}~\bibnamefont {Koning}},\ and\ \bibinfo {author} {\bibfnamefont {L.~A.}\ \bibnamefont {Hern{\'a}ndez}},\ }\bibfield  {title} {\bibinfo {title} {{Magnetic catalysis of charmonium in the vector channel}},\ }\href {https://doi.org/10.1103/6nz8-lwc1} {\bibfield  {journal} {\bibinfo  {journal} {Phys. Rev. D}\ }\textbf {\bibinfo {volume} {112}},\ \bibinfo {pages} {094049} (\bibinfo {year} {2025})},\ \Eprint {https://arxiv.org/abs/2510.09927} {arXiv:2510.09927 [hep-ph]} \BibitemShut {NoStop}%
\bibitem [{\citenamefont {Guo}\ \emph {et~al.}(2015)\citenamefont {Guo}, \citenamefont {Shi}, \citenamefont {Xu}, \citenamefont {Xu},\ and\ \citenamefont {Zhuang}}]{Guo:2015nsa}%
  \BibitemOpen
  \bibfield  {author} {\bibinfo {author} {\bibfnamefont {X.}~\bibnamefont {Guo}}, \bibinfo {author} {\bibfnamefont {S.}~\bibnamefont {Shi}}, \bibinfo {author} {\bibfnamefont {N.}~\bibnamefont {Xu}}, \bibinfo {author} {\bibfnamefont {Z.}~\bibnamefont {Xu}},\ and\ \bibinfo {author} {\bibfnamefont {P.}~\bibnamefont {Zhuang}},\ }\bibfield  {title} {\bibinfo {title} {{Magnetic field effect on charmonium formation in high energy nuclear collisions}},\ }\href {https://doi.org/10.1016/j.physletb.2015.10.038} {\bibfield  {journal} {\bibinfo  {journal} {Phys. Lett. B}\ }\textbf {\bibinfo {volume} {751}},\ \bibinfo {pages} {215} (\bibinfo {year} {2015})},\ \Eprint {https://arxiv.org/abs/1502.04407} {arXiv:1502.04407 [hep-ph]} \BibitemShut {NoStop}%
\bibitem [{\citenamefont {Suzuki}\ and\ \citenamefont {Lee}(2017)}]{Suzuki:2016fof}%
  \BibitemOpen
  \bibfield  {author} {\bibinfo {author} {\bibfnamefont {K.}~\bibnamefont {Suzuki}}\ and\ \bibinfo {author} {\bibfnamefont {S.~H.}\ \bibnamefont {Lee}},\ }\bibfield  {title} {\bibinfo {title} {{Delayed versus accelerated quarkonium formation in a magnetic field}},\ }\href {https://doi.org/10.1103/PhysRevC.96.035203} {\bibfield  {journal} {\bibinfo  {journal} {Phys. Rev. C}\ }\textbf {\bibinfo {volume} {96}},\ \bibinfo {pages} {035203} (\bibinfo {year} {2017})},\ \Eprint {https://arxiv.org/abs/1610.09853} {arXiv:1610.09853 [hep-ph]} \BibitemShut {NoStop}%
\bibitem [{\citenamefont {Dutta}\ and\ \citenamefont {Mazumder}(2018)}]{Dutta:2017pya}%
  \BibitemOpen
  \bibfield  {author} {\bibinfo {author} {\bibfnamefont {N.}~\bibnamefont {Dutta}}\ and\ \bibinfo {author} {\bibfnamefont {S.}~\bibnamefont {Mazumder}},\ }\bibfield  {title} {\bibinfo {title} {{Majorana flipping of quarkonium spin states in transient magnetic field}},\ }\href {https://doi.org/10.1140/epjc/s10052-018-6000-0} {\bibfield  {journal} {\bibinfo  {journal} {Eur. Phys. J. C}\ }\textbf {\bibinfo {volume} {78}},\ \bibinfo {pages} {525} (\bibinfo {year} {2018})},\ \Eprint {https://arxiv.org/abs/1704.04094} {arXiv:1704.04094 [nucl-th]} \BibitemShut {NoStop}%
\bibitem [{\citenamefont {Hoelck}\ and\ \citenamefont {Wolschin}(2017)}]{Hoelck:2017dby}%
  \BibitemOpen
  \bibfield  {author} {\bibinfo {author} {\bibfnamefont {J.}~\bibnamefont {Hoelck}}\ and\ \bibinfo {author} {\bibfnamefont {G.}~\bibnamefont {Wolschin}},\ }\bibfield  {title} {\bibinfo {title} {{Electromagnetic field effects on $\Upsilon$-meson dissociation in PbPb collisions at LHC energies}},\ }\href {https://doi.org/10.1140/epja/i2017-12441-0} {\bibfield  {journal} {\bibinfo  {journal} {Eur. Phys. J. A}\ }\textbf {\bibinfo {volume} {53}},\ \bibinfo {pages} {241} (\bibinfo {year} {2017})},\ \Eprint {https://arxiv.org/abs/1712.06871} {arXiv:1712.06871 [hep-ph]} \BibitemShut {NoStop}%
\bibitem [{\citenamefont {Bagchi}\ \emph {et~al.}(2023)\citenamefont {Bagchi}, \citenamefont {Dutta}, \citenamefont {Chatterjee},\ and\ \citenamefont {Adhya}}]{Bagchi:2018olp}%
  \BibitemOpen
  \bibfield  {author} {\bibinfo {author} {\bibfnamefont {P.}~\bibnamefont {Bagchi}}, \bibinfo {author} {\bibfnamefont {N.}~\bibnamefont {Dutta}}, \bibinfo {author} {\bibfnamefont {B.}~\bibnamefont {Chatterjee}},\ and\ \bibinfo {author} {\bibfnamefont {S.~P.}\ \bibnamefont {Adhya}},\ }\bibfield  {title} {\bibinfo {title} {{Dissociation of heavy quarkonium states in a rapidly varying strong magnetic field}},\ }\href {https://doi.org/10.1142/S0217732323500359} {\bibfield  {journal} {\bibinfo  {journal} {Mod. Phys. Lett. A}\ }\textbf {\bibinfo {volume} {38}},\ \bibinfo {pages} {2350035} (\bibinfo {year} {2023})},\ \Eprint {https://arxiv.org/abs/1805.04082} {arXiv:1805.04082 [nucl-th]} \BibitemShut {NoStop}%
\bibitem [{\citenamefont {Iwasaki}\ \emph {et~al.}(2021{\natexlab{b}})\citenamefont {Iwasaki}, \citenamefont {Jido}, \citenamefont {Oka},\ and\ \citenamefont {Suzuki}}]{Iwasaki:2021kms}%
  \BibitemOpen
  \bibfield  {author} {\bibinfo {author} {\bibfnamefont {S.}~\bibnamefont {Iwasaki}}, \bibinfo {author} {\bibfnamefont {D.}~\bibnamefont {Jido}}, \bibinfo {author} {\bibfnamefont {M.}~\bibnamefont {Oka}},\ and\ \bibinfo {author} {\bibfnamefont {K.}~\bibnamefont {Suzuki}},\ }\bibfield  {title} {\bibinfo {title} {{Survival probabilities of charmonia as a clue to measure transient magnetic fields}},\ }\href {https://doi.org/10.1016/j.physletb.2021.136498} {\bibfield  {journal} {\bibinfo  {journal} {Phys. Lett. B}\ }\textbf {\bibinfo {volume} {820}},\ \bibinfo {pages} {136498} (\bibinfo {year} {2021}{\natexlab{b}})},\ \Eprint {https://arxiv.org/abs/2104.13989} {arXiv:2104.13989 [hep-ph]} \BibitemShut {NoStop}%
\bibitem [{\citenamefont {Arifi}\ and\ \citenamefont {Suzuki}(2026)}]{Arifi:2025atv}%
  \BibitemOpen
  \bibfield  {author} {\bibinfo {author} {\bibfnamefont {A.~J.}\ \bibnamefont {Arifi}}\ and\ \bibinfo {author} {\bibfnamefont {K.}~\bibnamefont {Suzuki}},\ }\bibfield  {title} {\bibinfo {title} {{Landau-Zener-St{\"u}ckelberg-Majorana dynamics of magnetized quarkonia}},\ }\href {https://doi.org/10.1103/xc28-3v8r} {\bibfield  {journal} {\bibinfo  {journal} {Phys. Rev. D}\ }\textbf {\bibinfo {volume} {113}},\ \bibinfo {pages} {054047} (\bibinfo {year} {2026})},\ \Eprint {https://arxiv.org/abs/2512.24072} {arXiv:2512.24072 [hep-ph]} \BibitemShut {NoStop}%
\bibitem [{\citenamefont {Miransky}\ and\ \citenamefont {Shovkovy}(2002)}]{Miransky:2002rp}%
  \BibitemOpen
  \bibfield  {author} {\bibinfo {author} {\bibfnamefont {V.~A.}\ \bibnamefont {Miransky}}\ and\ \bibinfo {author} {\bibfnamefont {I.~A.}\ \bibnamefont {Shovkovy}},\ }\bibfield  {title} {\bibinfo {title} {{Magnetic catalysis and anisotropic confinement in QCD}},\ }\href {https://doi.org/10.1103/PhysRevD.66.045006} {\bibfield  {journal} {\bibinfo  {journal} {Phys. Rev. D}\ }\textbf {\bibinfo {volume} {66}},\ \bibinfo {pages} {045006} (\bibinfo {year} {2002})},\ \Eprint {https://arxiv.org/abs/hep-ph/0205348} {arXiv:hep-ph/0205348 [hep-ph]} \BibitemShut {NoStop}%
\bibitem [{\citenamefont {Chernodub}(2014)}]{Chernodub:2010xce}%
  \BibitemOpen
  \bibfield  {author} {\bibinfo {author} {\bibfnamefont {M.~N.}\ \bibnamefont {Chernodub}},\ }\bibfield  {title} {\bibinfo {title} {{QCD string breaking in strong magnetic field}},\ }\href {https://doi.org/10.1142/S0217732314501624} {\bibfield  {journal} {\bibinfo  {journal} {Mod. Phys. Lett. A}\ }\textbf {\bibinfo {volume} {29}},\ \bibinfo {pages} {1450162} (\bibinfo {year} {2014})},\ \Eprint {https://arxiv.org/abs/1001.0570} {arXiv:1001.0570 [hep-ph]} \BibitemShut {NoStop}%
\bibitem [{\citenamefont {Andreichikov}\ \emph {et~al.}(2013{\natexlab{a}})\citenamefont {Andreichikov}, \citenamefont {Orlovsky},\ and\ \citenamefont {Simonov}}]{Andreichikov:2012xe}%
  \BibitemOpen
  \bibfield  {author} {\bibinfo {author} {\bibfnamefont {M.~A.}\ \bibnamefont {Andreichikov}}, \bibinfo {author} {\bibfnamefont {V.~D.}\ \bibnamefont {Orlovsky}},\ and\ \bibinfo {author} {\bibfnamefont {{\relax Yu}.~A.}\ \bibnamefont {Simonov}},\ }\bibfield  {title} {\bibinfo {title} {{Asymptotic Freedom in Strong Magnetic Fields}},\ }\href {https://doi.org/10.1103/PhysRevLett.110.162002} {\bibfield  {journal} {\bibinfo  {journal} {Phys. Rev. Lett.}\ }\textbf {\bibinfo {volume} {110}},\ \bibinfo {pages} {162002} (\bibinfo {year} {2013}{\natexlab{a}})},\ \Eprint {https://arxiv.org/abs/1211.6568} {arXiv:1211.6568 [hep-ph]} \BibitemShut {NoStop}%
\bibitem [{\citenamefont {Simonov}\ and\ \citenamefont {Trusov}(2015)}]{Simonov:2015yka}%
  \BibitemOpen
  \bibfield  {author} {\bibinfo {author} {\bibfnamefont {Y.~A.}\ \bibnamefont {Simonov}}\ and\ \bibinfo {author} {\bibfnamefont {M.~A.}\ \bibnamefont {Trusov}},\ }\bibfield  {title} {\bibinfo {title} {{Confinement and $\alpha_s$ in a strong magnetic field}},\ }\href {https://doi.org/10.1016/j.physletb.2015.05.032} {\bibfield  {journal} {\bibinfo  {journal} {Phys. Lett. B}\ }\textbf {\bibinfo {volume} {747}},\ \bibinfo {pages} {48} (\bibinfo {year} {2015})},\ \Eprint {https://arxiv.org/abs/1503.08531} {arXiv:1503.08531 [hep-ph]} \BibitemShut {NoStop}%
\bibitem [{\citenamefont {Bohra}\ \emph {et~al.}(2020)\citenamefont {Bohra}, \citenamefont {Dudal}, \citenamefont {Hajilou},\ and\ \citenamefont {Mahapatra}}]{Bohra:2019ebj}%
  \BibitemOpen
  \bibfield  {author} {\bibinfo {author} {\bibfnamefont {H.}~\bibnamefont {Bohra}}, \bibinfo {author} {\bibfnamefont {D.}~\bibnamefont {Dudal}}, \bibinfo {author} {\bibfnamefont {A.}~\bibnamefont {Hajilou}},\ and\ \bibinfo {author} {\bibfnamefont {S.}~\bibnamefont {Mahapatra}},\ }\bibfield  {title} {\bibinfo {title} {{Anisotropic string tensions and inversely magnetic catalyzed deconfinement from a dynamical AdS/QCD model}},\ }\href {https://doi.org/10.1016/j.physletb.2019.135184} {\bibfield  {journal} {\bibinfo  {journal} {Phys. Lett. B}\ }\textbf {\bibinfo {volume} {801}},\ \bibinfo {pages} {135184} (\bibinfo {year} {2020})},\ \Eprint {https://arxiv.org/abs/1907.01852} {arXiv:1907.01852 [hep-th]} \BibitemShut {NoStop}%
\bibitem [{\citenamefont {G{\"u}rsoy}\ \emph {et~al.}(2021)\citenamefont {G{\"u}rsoy}, \citenamefont {J{\"a}rvinen}, \citenamefont {Nijs},\ and\ \citenamefont {Pedraza}}]{Gursoy:2020kjd}%
  \BibitemOpen
  \bibfield  {author} {\bibinfo {author} {\bibfnamefont {U.}~\bibnamefont {G{\"u}rsoy}}, \bibinfo {author} {\bibfnamefont {M.}~\bibnamefont {J{\"a}rvinen}}, \bibinfo {author} {\bibfnamefont {G.}~\bibnamefont {Nijs}},\ and\ \bibinfo {author} {\bibfnamefont {J.~F.}\ \bibnamefont {Pedraza}},\ }\bibfield  {title} {\bibinfo {title} {{On the interplay between magnetic field and anisotropy in holographic QCD}},\ }\href {https://doi.org/10.1007/JHEP03(2021)180} {\bibfield  {journal} {\bibinfo  {journal} {JHEP}\ }\textbf {\bibinfo {volume} {03}},\ \bibinfo {pages} {(2021) 180}},\ \Eprint {https://arxiv.org/abs/2011.09474} {arXiv:2011.09474 [hep-th]} \BibitemShut {NoStop}%
\bibitem [{\citenamefont {Aref'eva}\ \emph {et~al.}(2020)\citenamefont {Aref'eva}, \citenamefont {Rannu},\ and\ \citenamefont {Slepov}}]{Arefeva:2020bjk}%
  \BibitemOpen
  \bibfield  {author} {\bibinfo {author} {\bibfnamefont {I.~Y.}\ \bibnamefont {Aref'eva}}, \bibinfo {author} {\bibfnamefont {K.}~\bibnamefont {Rannu}},\ and\ \bibinfo {author} {\bibfnamefont {P.}~\bibnamefont {Slepov}},\ }\bibfield  {title} {\bibinfo {title} {{Energy Loss in Holographic Anisotropic Model for Heavy Quarks in External Magnetic Field}},\ }\href@noop {} {\  (\bibinfo {year} {2020})},\ \Eprint {https://arxiv.org/abs/2012.05758} {arXiv:2012.05758 [hep-th]} \BibitemShut {NoStop}%
\bibitem [{\citenamefont {Aref'eva}\ \emph {et~al.}(2026)\citenamefont {Aref'eva}, \citenamefont {Hajilou}, \citenamefont {Rannu},\ and\ \citenamefont {Slepov}}]{Arefeva:2026yms}%
  \BibitemOpen
  \bibfield  {author} {\bibinfo {author} {\bibfnamefont {I.~Y.}\ \bibnamefont {Aref'eva}}, \bibinfo {author} {\bibfnamefont {A.}~\bibnamefont {Hajilou}}, \bibinfo {author} {\bibfnamefont {K.}~\bibnamefont {Rannu}},\ and\ \bibinfo {author} {\bibfnamefont {P.}~\bibnamefont {Slepov}},\ }\bibfield  {title} {\bibinfo {title} {{Spatial Wilson loops and energy loss for heavy quarks in a magnetized holographic QCD model}},\ }\href {https://doi.org/10.1103/n337-n4l3} {\bibfield  {journal} {\bibinfo  {journal} {Phys. Rev. D}\ }\textbf {\bibinfo {volume} {113}},\ \bibinfo {pages} {106004} (\bibinfo {year} {2026})},\ \Eprint {https://arxiv.org/abs/2601.09611} {arXiv:2601.09611 [hep-th]} \BibitemShut {NoStop}%
\bibitem [{\citenamefont {Barnes}\ \emph {et~al.}(2005)\citenamefont {Barnes}, \citenamefont {Godfrey},\ and\ \citenamefont {Swanson}}]{Barnes:2005pb}%
  \BibitemOpen
  \bibfield  {author} {\bibinfo {author} {\bibfnamefont {T.}~\bibnamefont {Barnes}}, \bibinfo {author} {\bibfnamefont {S.}~\bibnamefont {Godfrey}},\ and\ \bibinfo {author} {\bibfnamefont {E.~S.}\ \bibnamefont {Swanson}},\ }\bibfield  {title} {\bibinfo {title} {{Higher charmonia}},\ }\href {https://doi.org/10.1103/PhysRevD.72.054026} {\bibfield  {journal} {\bibinfo  {journal} {Phys. Rev. D}\ }\textbf {\bibinfo {volume} {72}},\ \bibinfo {pages} {054026} (\bibinfo {year} {2005})},\ \Eprint {https://arxiv.org/abs/hep-ph/0505002} {arXiv:hep-ph/0505002 [hep-ph]} \BibitemShut {NoStop}%
\bibitem [{\citenamefont {Kawanai}\ and\ \citenamefont {Sasaki}(2015)}]{Kawanai:2015tga}%
  \BibitemOpen
  \bibfield  {author} {\bibinfo {author} {\bibfnamefont {T.}~\bibnamefont {Kawanai}}\ and\ \bibinfo {author} {\bibfnamefont {S.}~\bibnamefont {Sasaki}},\ }\bibfield  {title} {\bibinfo {title} {{Potential description of charmonium and charmed-strange mesons from lattice QCD}},\ }\href {https://doi.org/10.1103/PhysRevD.92.094503} {\bibfield  {journal} {\bibinfo  {journal} {Phys. Rev. D}\ }\textbf {\bibinfo {volume} {92}},\ \bibinfo {pages} {094503} (\bibinfo {year} {2015})},\ \Eprint {https://arxiv.org/abs/1508.02178} {arXiv:1508.02178 [hep-lat]} \BibitemShut {NoStop}%
\bibitem [{\citenamefont {Kawanai}\ and\ \citenamefont {Sasaki}(2012)}]{Kawanai:2011jt}%
  \BibitemOpen
  \bibfield  {author} {\bibinfo {author} {\bibfnamefont {T.}~\bibnamefont {Kawanai}}\ and\ \bibinfo {author} {\bibfnamefont {S.}~\bibnamefont {Sasaki}},\ }\bibfield  {title} {\bibinfo {title} {{Charmonium potential from full lattice QCD}},\ }\href {https://doi.org/10.1103/PhysRevD.85.091503} {\bibfield  {journal} {\bibinfo  {journal} {Phys. Rev. D}\ }\textbf {\bibinfo {volume} {85}},\ \bibinfo {pages} {091503} (\bibinfo {year} {2012})},\ \Eprint {https://arxiv.org/abs/1110.0888} {arXiv:1110.0888 [hep-lat]} \BibitemShut {NoStop}%
\bibitem [{\citenamefont {Johnson}\ and\ \citenamefont {Lippmann}(1949)}]{Johnson:1949}%
  \BibitemOpen
  \bibfield  {author} {\bibinfo {author} {\bibfnamefont {M.~H.}\ \bibnamefont {Johnson}}\ and\ \bibinfo {author} {\bibfnamefont {B.~A.}\ \bibnamefont {Lippmann}},\ }\bibfield  {title} {\bibinfo {title} {Motion in a constant magnetic field},\ }\href {https://doi.org/10.1103/PhysRev.76.828} {\bibfield  {journal} {\bibinfo  {journal} {Phys. Rev.}\ }\textbf {\bibinfo {volume} {76}},\ \bibinfo {pages} {828} (\bibinfo {year} {1949})}\BibitemShut {NoStop}%
\bibitem [{\citenamefont {Andreichikov}\ \emph {et~al.}(2013{\natexlab{b}})\citenamefont {Andreichikov}, \citenamefont {Kerbikov}, \citenamefont {Orlovsky},\ and\ \citenamefont {Simonov}}]{Andreichikov:2013zba}%
  \BibitemOpen
  \bibfield  {author} {\bibinfo {author} {\bibfnamefont {M.~A.}\ \bibnamefont {Andreichikov}}, \bibinfo {author} {\bibfnamefont {B.~O.}\ \bibnamefont {Kerbikov}}, \bibinfo {author} {\bibfnamefont {V.~D.}\ \bibnamefont {Orlovsky}},\ and\ \bibinfo {author} {\bibfnamefont {Y.~A.}\ \bibnamefont {Simonov}},\ }\bibfield  {title} {\bibinfo {title} {{Meson spectrum in strong magnetic fields}},\ }\href {https://doi.org/10.1103/PhysRevD.87.094029} {\bibfield  {journal} {\bibinfo  {journal} {Phys. Rev. D}\ }\textbf {\bibinfo {volume} {87}},\ \bibinfo {pages} {094029} (\bibinfo {year} {2013}{\natexlab{b}})},\ \Eprint {https://arxiv.org/abs/1304.2533} {arXiv:1304.2533 [hep-ph]} \BibitemShut {NoStop}%
\bibitem [{\citenamefont {Kamimura}(1988)}]{Kamimura:1988zz}%
  \BibitemOpen
  \bibfield  {author} {\bibinfo {author} {\bibfnamefont {M.}~\bibnamefont {Kamimura}},\ }\bibfield  {title} {\bibinfo {title} {{Nonadiabatic coupled-rearrangement-channel approach to muonic molecules}},\ }\href {https://doi.org/10.1103/PhysRevA.38.621} {\bibfield  {journal} {\bibinfo  {journal} {Phys. Rev. A}\ }\textbf {\bibinfo {volume} {38}},\ \bibinfo {pages} {621} (\bibinfo {year} {1988})}\BibitemShut {NoStop}%
\bibitem [{\citenamefont {Hiyama}\ \emph {et~al.}(2003)\citenamefont {Hiyama}, \citenamefont {Kino},\ and\ \citenamefont {Kamimura}}]{Hiyama:2003cu}%
  \BibitemOpen
  \bibfield  {author} {\bibinfo {author} {\bibfnamefont {E.}~\bibnamefont {Hiyama}}, \bibinfo {author} {\bibfnamefont {Y.}~\bibnamefont {Kino}},\ and\ \bibinfo {author} {\bibfnamefont {M.}~\bibnamefont {Kamimura}},\ }\bibfield  {title} {\bibinfo {title} {Gaussian expansion method for few-body systems},\ }\href {https://doi.org/10.1016/S0146-6410(03)90015-9} {\bibfield  {journal} {\bibinfo  {journal} {Prog. Part. Nucl. Phys.}\ }\textbf {\bibinfo {volume} {51}},\ \bibinfo {pages} {223} (\bibinfo {year} {2003})}\BibitemShut {NoStop}%
\bibitem [{Note1()}]{Note1}%
  \BibitemOpen
  \bibinfo {note} {We note that the transverse-sector parameters exhibit strong correlations, particularly between $a_T$ and $\kappa _T$, since the asymptotic region is less constrained, whereas the longitudinal-sector parameters are only weakly correlated because we impose $a_L=0$.}\BibitemShut {Stop}%
\bibitem [{\citenamefont {Orlovsky}\ and\ \citenamefont {Simonov}(2013)}]{Orlovsky:2013gha}%
  \BibitemOpen
  \bibfield  {author} {\bibinfo {author} {\bibfnamefont {V.~D.}\ \bibnamefont {Orlovsky}}\ and\ \bibinfo {author} {\bibfnamefont {Y.~A.}\ \bibnamefont {Simonov}},\ }\bibfield  {title} {\bibinfo {title} {{Nambu-Goldstone mesons in strong magnetic field}},\ }\href {https://doi.org/10.1007/JHEP09(2013)136} {\bibfield  {journal} {\bibinfo  {journal} {JHEP}\ }\textbf {\bibinfo {volume} {09}},\ \bibinfo {pages} {(2013) 136}},\ \Eprint {https://arxiv.org/abs/1306.2232} {arXiv:1306.2232 [hep-ph]} \BibitemShut {NoStop}%
\bibitem [{\citenamefont {Taya}(2015)}]{Taya:2014nha}%
  \BibitemOpen
  \bibfield  {author} {\bibinfo {author} {\bibfnamefont {H.}~\bibnamefont {Taya}},\ }\bibfield  {title} {\bibinfo {title} {{Hadron Masses in Strong Magnetic Fields}},\ }\href {https://doi.org/10.1103/PhysRevD.92.014038} {\bibfield  {journal} {\bibinfo  {journal} {Phys. Rev. D}\ }\textbf {\bibinfo {volume} {92}},\ \bibinfo {pages} {014038} (\bibinfo {year} {2015})},\ \Eprint {https://arxiv.org/abs/1412.6877} {arXiv:1412.6877 [hep-ph]} \BibitemShut {NoStop}%
\bibitem [{\citenamefont {Andreichikov}\ \emph {et~al.}(2017)\citenamefont {Andreichikov}, \citenamefont {Kerbikov}, \citenamefont {Luschevskaya}, \citenamefont {Simonov},\ and\ \citenamefont {Solovjeva}}]{Andreichikov:2016ayj}%
  \BibitemOpen
  \bibfield  {author} {\bibinfo {author} {\bibfnamefont {M.~A.}\ \bibnamefont {Andreichikov}}, \bibinfo {author} {\bibfnamefont {B.~O.}\ \bibnamefont {Kerbikov}}, \bibinfo {author} {\bibfnamefont {E.~V.}\ \bibnamefont {Luschevskaya}}, \bibinfo {author} {\bibfnamefont {{\relax Yu}.~A.}\ \bibnamefont {Simonov}},\ and\ \bibinfo {author} {\bibfnamefont {O.~E.}\ \bibnamefont {Solovjeva}},\ }\bibfield  {title} {\bibinfo {title} {{The Evolution of Meson Masses in a Strong Magnetic Field}},\ }\href {https://doi.org/10.1007/JHEP05(2017)007} {\bibfield  {journal} {\bibinfo  {journal} {JHEP}\ }\textbf {\bibinfo {volume} {05}},\ \bibinfo {pages} {(2017) 007}},\ \Eprint {https://arxiv.org/abs/1610.06887} {arXiv:1610.06887 [hep-ph]} \BibitemShut {NoStop}%
\bibitem [{\citenamefont {Kojo}(2021)}]{Kojo:2021gvm}%
  \BibitemOpen
  \bibfield  {author} {\bibinfo {author} {\bibfnamefont {T.}~\bibnamefont {Kojo}},\ }\bibfield  {title} {\bibinfo {title} {{Neutral and charged mesons in magnetic fields: A resonance gas in a non-relativistic quark model}},\ }\href {https://doi.org/10.1140/epja/s10050-021-00629-y} {\bibfield  {journal} {\bibinfo  {journal} {Eur. Phys. J. A}\ }\textbf {\bibinfo {volume} {57}},\ \bibinfo {pages} {317} (\bibinfo {year} {2021})},\ \Eprint {https://arxiv.org/abs/2104.00376} {arXiv:2104.00376 [hep-ph]} \BibitemShut {NoStop}%
\bibitem [{\citenamefont {Rabi}(1928)}]{Rabi:1928}%
  \BibitemOpen
  \bibfield  {author} {\bibinfo {author} {\bibfnamefont {I.~I.}\ \bibnamefont {Rabi}},\ }\bibfield  {title} {\bibinfo {title} {{Das freie Elektron im homogenen Magnetfeld nach der Diracschen Theorie}},\ }\href {https://doi.org/10.1007/BF01333634} {\bibfield  {journal} {\bibinfo  {journal} {Z. Phys.}\ }\textbf {\bibinfo {volume} {49}},\ \bibinfo {pages} {507} (\bibinfo {year} {1928})}\BibitemShut {NoStop}%
\end{thebibliography}%

\end{document}